\documentclass[twocolumn,pre,floatfix]{revtex4}
\usepackage{psfrag,epsfig,amsfonts,amssymb,amsmath,wasysym,bm}
\usepackage{dcolumn}
\usepackage{bbold}
\usepackage[normalem]{ulem}
\usepackage{color}

\usepackage[inline]{enumitem} % enumerated lists
\usepackage{hyperref}
\newcommand{\ket}[1]{\lvert #1 \rangle} 	% ket
\newcommand{\bra}[1]{\langle #1 \rvert}	% bra
\newcommand{\ketN}[1]{\lvert #1 \rangle_{\! 0}} 	% ket
\newcommand{\braN}[1]{{_0\!}\langle #1 \rvert}	% bra
\newcommand{\sUp}{\uparrow}		% spin up
\newcommand{\sDn}{\downarrow}	% spin down

\newcommand{\<}{\left\langle}	% angular brackets
\renewcommand{\>}{\right\rangle}	% "

\newcommand{\e}{\mathrm{e}}		% Euler number
\newcommand{\I}{\mathrm{i}}		% imaginary unit
\newcommand{\lvsp}{\varepsilon}

\newcommand{\id}{\mathbb 1}
\newcommand{\RR}{{\mathbb R}}
\newcommand{\CC}{{\mathbb C}}

\newcommand{\tr}{\mbox{Tr}}

\providecommand{\av}[1]{[#1]_V}

\newcommand{\lmat}{\left( \begin{matrix}}	% matrix begin
\newcommand{\rmat}{\end{matrix} \right)}	% matrix end
\newcommand{\CFK}{%
  \operatornamewithlimits{%
    \mathchoice
     {\vcenter{\hbox{\huge $\mathcal{K}$}}}
     {\vcenter{\hbox{\Large $\mathcal{K}$}}}
     {\mathcal{K}}
     {\mathcal{K}}}
}

\begin{document}

\title{Typical relaxation of perturbed quantum many-body systems}

\author{Lennart Dabelow}
\email{ldabelow@physik.uni-bielefeld.de}
\author{Peter Reimann}
\email{reimann@physik.uni-bielefeld.de}
\affiliation{Fakult\"at f\"ur Physik, 
Universit\"at Bielefeld, 
33615 Bielefeld, Germany}
\date{\today}

\begin{abstract}
We substantially extend our relaxation theory for perturbed many-body quantum systems 
from [Phys.~Rev.~Lett.~{\bf 124}, 120602 (2020)] by establishing an analytical prediction 
for the time-dependent observable expectation values which depends on only two 
characteristic parameters of the perturbation operator:
its overall strength and its range or band width.
Compared to the previous theory,
a significantly larger range of perturbation strengths is covered.
The results are obtained within a typicality framework by solving the pertinent random 
matrix problem exactly for a certain class of banded perturbations and by demonstrating 
the (approximative) universality of these solutions,
which allows us to adopt
them to considerably more general classes of perturbations.
We also verify the prediction by comparison with several numerical examples.  
\end{abstract}

\maketitle

%%%%%%%%%%%%%%%%%%%%%%%%%%%%%%%%%%%%%%%%%%%%%%%%%%%%
\section{Introduction}
\label{s1}

The question of how the behavior of a given system changes in 
response to a weak perturbation is ubiquitous in physics.
If many degrees of freedom are involved,
the microscopic dynamics is 
commonly expected to 
be extremely sensitive against small changes 
(chaotic \cite{haa10}),
so that
it is virtually impossible to theoretically
predict the response exactly or in terms of 
well-controlled approximations \cite{kam71}.
Yet, the actually observed behavior
in experiments and numerical simulations 
is often found to obey relatively simple 
and robust ``laws''.
Here, we specifically ask
how the temporal relaxation of an isolated 
many-body quantum system is altered by a small
modification of the Hamiltonian, and we
bridge the inevitable gap between what is 
theoretically feasible and what is actually 
observed by adopting
the general framework of
random matrix theory
\cite{haa10,bro81}.

As our starting point we utilize the
tools and results which we 
previously established in Ref.~\cite{dab20}
and summarize here in Sec.~\ref{s2}
(see also \cite{deu91,gen13,nat19} for some
related earlier works and Sec.~\ref{s5} for a more detailed discussion of their connection to our present approach):
Essentially, the idea is to consider not
one specific but rather an entire ensemble of 
perturbations, most of which still closely 
resemble the one of actual interest.
The first main result obtained in Ref.~\cite{dab20} was
an analytical prediction for the ensemble-averaged,
time-dependent deviations of the perturbed from the
unperturbed expectation values
of generic observables.
In a second step, it was shown that
nearly all members of the ensemble behave 
very similarly to the average.
Finally, it was argued that also the system 
of actual interest belongs to that vast majority. 
In the context of random matrix 
theory, this is a well-established 
line of reasoning, which has to our 
knowledge never been rigorously justified, 
but is extremely successful in practice \cite{bro81}.
In fact, it has been originally devised by Wigner 
for the very purpose of exploring  
chaotic quantum many-body systems
and is by now widely recognized as a 
remarkably effective tool in this context
\cite{haa10}.
We emphasize once more that such an approach 
should {\em not} be viewed as a randomization of 
the real physical perturbation \cite{gol10a,gol10b}.
Rather, the basic assertion is that
the ``true'' (non-random) perturbation 
belongs to the vast majority of all the very 
similarly behaving members of some
properly chosen ensemble.

Technically speaking, a key role in the above described
approach from Ref. \cite{dab20} is played by a non-linear 
integral equation, which so far could only be tackled
analytically in two limiting cases.
At the focus of our present paper is a much more detailed
analytical investigation of this non-linear integral equation
(Sec.~\ref{s3}).

The central results of this work are
summarized in Sec.~\ref{sMainResults} and established in detail in Sec.~\ref{s4}.
Our first main result is that a very large class of
perturbations can be extremely well characterized
by only two parameters.
One of them essentially describes the perturbation 
strength.
The other one quantifies the range of the 
perturbation
in terms of how quickly its matrix elements in the unperturbed eigenbasis decay with their energy difference.
Our second main result is a very general analytical 
approximation of the perturbed relaxation in terms of 
these two parameters.
As a validation of the adopted random matrix approach itself 
and of our analytical approximations within such an approach, 
we finally compare our predictions with numerically obtained 
results for various specific models (Sec.~\ref{s6}).
Altogether, notable analytical progress is thus achieved
with regard to the general topics of equilibration and thermalization
in isolated many-body quantum systems, which are attracting 
increasing theoretical and experimental interest during recent 
years, as reviewed, e.g., in Refs.
 \cite{gog16,dal16,mor18,bor16,nan15,lan16,tas16}.

%%%%%%%%%%%%%%%%%%%%%%%%%%%%%%%%%%%%%%%%%%%%%%%%%%%%
%%%%%%%%%%%%%%%%%%%%%%%%%%%%%%%%%%%%%%%%%%%%%%%%%%%%
\section{Setting the stage}
\label{s2}
In this section, we 
outline
the general
framework and the main results from the 
pertinent
predecessor
work \cite{dab20},
which in turn is very similar in its general spirit to
the hallmark paper by Deutsch \cite{deu91}
and its further developments for instance in Refs.
\cite{gen13,nat19,rei15,nat18}.

As announced in the Introduction, we 
study the temporal relaxation of an isolated 
many-body quantum system with
Hamiltonian
\begin{eqnarray}
H = H_0 + \lambda V
\ ,
\label{1}
\end{eqnarray}
where $H_0$ describes the unperturbed system,
$V$ the perturbation, and $\lambda$ the coupling.
At time $t=0$, the system is prepared in some pure or mixed, 
and generally far-from-equilibrium
initial state $\rho(0)$.
Considering the unperturbed relaxation behavior 
as given (known), our aim is to draw conclusions
about the time evolution of the same initial state 
when the dynamics is subject to sufficiently weak 
perturbations.

For example, $H_0$ may model two isolated subsystems
(or a system and its environment), and $\lambda V$ their interaction. 
If both are prepared in thermal equilibrium states with different
temperatures, the perturbation
causes a relaxation towards a new thermal 
equilibrium of the compound system.
More generally, already the unperturbed subsystems
may exhibit some non-trivial relaxation, which is
then modified by the perturbation.

Second, if the considered observable 
commutes with $H_0$ (constant of motion), 
one may ask for the response to a perturbation 
which breaks the corresponding symmetry.
Similarly, the initial state may 
commute with $H_0$ (steady state),
but not any more with $H$.

Third, analytical solutions may be available 
for $H_0$ but not for $H$.
For instance, $H_0$ may describe a 
non-interacting many-body system 
and $V$ the interactions,
or $H_0$ may be integrable 
and $H$ non-integrable.

Finally, interesting examples even without 
a steady long-time limit of the unperturbed 
system are conceivable and will be covered 
by our approach.

Focusing on possibly large but finite systems, 
the Hamiltonian $H$ in (\ref{1})
exhibits a discrete set of eigenvalues
$E_n$ and eigenvectors 
$\ket{n}$, with $n$ running from 
one to infinity, or, 
for instance for a spin model,
to some large but finite upper limit.
The initial state $\rho(0)$ evolves according to the Schr\"odinger 
or von Neumann equation, so that the state at a later time $t>0$ 
is given by $\rho(t) = e^{-i H t} \, \rho(0) \, e^{i H t}$
($\hbar = 1$), and the expectation value of 
an observable (self-adjoint operator) 
$A$ by
\begin{eqnarray}
\label{2}
	\< A \>_{\! \rho(t)} := \tr\{\rho(t) A\}=\sum_{m, n} e^{i (E_n - E_m) t} \, \rho_{mn}(0) \, A_{nm} \,,
\end{eqnarray}
where $\rho_{mn}(t) := \bra{m} \rho(t) \ket{n}$ and 
$A_{nm} := \bra{n} A \ket{m}$.
Likewise, the corresponding 
expectation values 
when the same initial state 
$\rho(0)$ evolves according to the
unperturbed Hamiltonian $H_0$
with eigenvalues $E^0_n$ and eigenvectors 
$\ketN{n}$ take the form
\begin{eqnarray}
\label{3}
\< A \>_{\!\rho_0\!(t)} := \tr\{\rho_0(t) A\} =\sum_{m, n} e^{i (E^0_n - E^0_m) t} \, \rho^0_{mn}(0) \, A^0_{nm} 
\end{eqnarray}
with $\rho^0_{mn}(0) := \braN{m} \rho(0) \ketN{n}$
and $A^0_{nm} := \braN{n} A \ketN{m}$.

As usual in this context \cite{gog16,dal16,mor18},
we take for granted that only energies $E_{n}$ 
within some 
sufficiently small
energy interval $I_{\! E}$ 
entail non-negligible level populations
$\rho_{nn}(0)$ so that the
(locally averaged) mean level spacing 
$\lvsp$ is approximately constant 
throughout $I_{\! E}$, and likewise for the 
unperturbed populations $\rho^0_{nn}(0)$
\cite{deu91}.
This assumption of an approximately constant density of states $\varepsilon^{-1}$ for both $H_0$ and $H$ in particular requires that the perturbation should not become so strong that it modifies the system's thermodynamic properties since the density of states is directly related to (the derivative of) Boltzmann's entropy.

As shown in \cite{dab20}, we thus can
replace both $E_n-E_m$ in (\ref{2}) and $E^0_n-E^0_m$ in (\ref{3}) 
by $(n-m)\lvsp$ in extremely good approximation.
Finally, one can employ the overlaps of the 
perturbed and unperturbed energy eigenstates,
\begin{equation}
	U_{mn} := \langle m | n \rangle_{\!0}
	\,,
\label{4}
\end{equation}
to rewrite also the right hand side of (\ref{2})
in terms of the unperturbed matrix elements
$\rho^0_{mn}(0)$ and $A^0_{nm}$
(see also Sec.~\ref{s51}).
Hence, the comparison between the 
perturbed and unperturbed relaxation behavior
in (\ref{2}) and (\ref{3})
essentially boils down to gathering sufficient information 
about the unitary basis transformation in (\ref{4}).
The simplest idea that immediately comes
to mind is to utilize in (\ref{1}) 
elementary (Rayleigh-Schr\"odinger) 
perturbation theory in order to determine 
those matrix elements in (\ref{4}).
However, such an approach is limited
to exceedingly small $\lambda$ values 
in (\ref{1}) due the extremely large level
density $\varepsilon^{-1}$ and the concomitant 
small denominators in such a perturbative
treatment.
On the one hand, such exceedingly weak perturbations will hardly affect the expectation values in~\eqref{2} on reasonable time scales at all.
For sufficiently strong perturbations, on the other hand,
a perturbative expansion of the propagator will generally be limited to uninterestingly small time scales compared to the relevant relaxation time.
In other words, some alternative,
non-perturbative approach is indispensable.

We thus adopted in \cite{dab20} the well-established
approach \cite{deu91,gen13,nat19,rei15,nat18}
to temporarily consider an entire statistical 
ensemble of similar perturbations instead 
of one particular $V$ in (\ref{1})
(see also Introduction and Sec.~\ref{s6}
for the general conceptual
ideas behind such an approach).
More precisely speaking, 
expressing the operator $V$ in the eigenbasis 
of the unperturbed system, 
we choose an ensemble of matrices
$V^0_{mn} := \braN{m} V \ketN{n}$
whose statistics still reflects the essential 
properties of the ``true'' perturbation $V$
in (\ref{1}) as closely as possible.
For instance, the $V^0_{mn}$ 
may (but need not) exhibit a so-called 
banded and/or sparse matrix structure.
Technically speaking, our main assumptions regarding
the admitted random matrix ensembles are that the matrix
elements $V^0_{mn}$ must be statistically independent 
(apart from $V^0_{nm} = (V^0_{mn})^\ast$) and of zero 
average (unbiased).
Moreover, the second moments 
(or variances)
should only depend
-- at least within the relevant energy interval $I_{\! E}$ 
and in sufficiently good approximation --
on the energy difference $E^0_m-E^0_n$ of the
two levels $\ketN{m}$, $\ketN{n}$ which are
coupled via $V^0_{mn}$.
Recalling that $E^0_m - E^0_n\simeq (m-n)\lvsp$
one might equally well say that the statistical
properties (or at least the variances) 
of the $V^0_{mn}$
do not depend separately on $m$ and $n$, 
but only on the difference $m-n$.
Denoting averages over the ensemble of perturbations 
by $\av{\,\cdots}$,
those second moments can thus 
be (approximately) represented as
\begin{equation}
	\av{\lvert V^0_{mn} \rvert^2} = \sigma^2(E^0_m - E^0_n)
\label{5}
\end{equation}
for some suitable ``variance function''
or \emph{perturbation profile} $\sigma^2(x)$,
which furthermore must satisfy
\begin{eqnarray}
\sigma^2(-x)=\sigma^2(x)
\label{6}
\end{eqnarray}
due to $V^0_{nm} = (V^0_{mn})^\ast$.
Specifically, in cases where $V^0_{mn}$ exhibits a banded
matrix structure (see above), the variances $\sigma^2(x)$ in (\ref{5})
will approach zero for large energy differences $x$.
Regarding the true perturbation of interest,
the key assumption is thus that it exhibits a well-defined perturbation profile in the sense that Eq.~\eqref{5} holds when replacing the ensemble average $\av{ \,\cdots }$ by a ``local'' (running) average over close-by energy levels.
As a result, $\sigma^2(x)$ is essentially a smooth function by construction.
Moreover, the following typicality argument can only be expected to include the true $V$ if the principle mechanism responsible for the dynamical modifications induced by the perturbation is captured by the perturbation profile as the defining property of the considered ensemble
(see also the beginning of Sec.~\ref{s6}).

Within this general framework, the main result obtained
in Ref. \cite{dab20} is that,
for the overwhelming majority of $V$'s in the random matrix 
ensembles specified above, the perturbed relaxation
from (\ref{2}) will be very well approximated by
\begin{equation}
	\< A \>_{\!\rho(t)}
		=  \< A \>_{\!\bar\rho}  
		% \Ath
		+ \lvert g(t) \rvert^2 \left\{ \< A \>_{\!\rho_0(t)} -  \< A \>_{\!\bar\rho}  \right\}
\label{7}
\end{equation}
for all $\lambda$ admitted below Eq.~\eqref{3}.
Here $\< A \>_{\!\rho_0(t)}$ is the unperturbed behavior 
from (\ref{3}) and
$\bar\rho$ is the time-averaged state obtained from $\rho(t)$,
i.e., the so-called diagonal ensemble associated with $\rho(0)$ and the perturbed Hamiltonian $H$ from~\eqref{1}
\cite{gog16,dal16,mor18}.
Generically, $\< A \>_{\!\bar\rho}$ is thus expected to coincide with the pertinent thermal value \cite{deu91, rei15, nat18, dab20}.
Finally, the function $g(t)$ in (\ref{7}) is defined as
\begin{eqnarray}
g(t):= \frac{1}{\pi}\,\lim_{\eta\downarrow 0}\int dE\, e^{iEt}\, \mbox{Im}\,G(E-i\eta) 
\ ,
\label{8}
\end{eqnarray}
where $G(z)$
solves
the non-linear integral equation 
\begin{eqnarray}
G(z)\,\left[z-\frac{\lambda^2}{\lvsp}\int dx\, G(z-x) \, \sigma^2(x)\right] = 1 \ .
\label{9}
\end{eqnarray}
The connection to the perturbed Hamiltonian from~\eqref{1} is established by observing that the function $G(z)$ encodes the ensemble average of the resolvent or Green's function $\mathcal{G}(z) := (z-H)^{-1}$
via $G(z-H_0) = \av{ \mathcal{G}(z) }$.
In particular, the poles of $\mathcal G(z)$ correspond to the eigenvalues $E_n$ of $H$, whereas the matrix elements $\braN{m} \mathcal{G}(z) \ketN{n}$ in the vicinity of these poles are related to the overlaps between the perturbed and unperturbed eigenvectors from~\eqref{4},
see also Appendix \ref{app1} and the Supplemental Material of Ref.~\cite{dab20} for some additional details.

Overall,
these findings~\eqref{7}--\eqref{9} are 
essentially (apart from some minor technical subtleties)
understood to be asymptotically exact for systems with 
many degrees of freedom, i.e.,
for large but finite systems there will be
some quantitatively unknown subleading 
order corrections \cite{dab20}.
In particular, the probability to randomly sample
a member $V$ of the ensemble which exhibits notable 
deviations from (\ref{7}) is exponentially small in the
system's degrees of freedom \cite{dab20}.
More precisely speaking, this probability decreases with the effective number of unperturbed energy levels contributing to any perturbed eigenvector via~\eqref{4}.
In principle, the theory may thus equally apply to systems with fewer degrees of freedom provided that they exhibit a sufficiently dense energy spectrum and sufficiently strong mixing of energy levels by the perturbation such that all formal requirements from above are still fulfilled.

Tackling (\ref{9}) and then (\ref{8}) by analytical means is
the main objective of our present paper.

%%%%%%%%%%%%%%%%%%%%%%%%%%%%%%%%%%%%%%%%%%%%%%%%%%%%
%%%%%%%%%%%%%%%%%%%%%%%%%%%%%%%%%%%%%%%%%%%%%%%%%%%%
\section{Main results}
\label{sMainResults}

Before going into the detailed derivations, we briefly summarize the main new findings of our present paper.
These consist in relatively simple but expressive representations for the function $g(t)$, which describes how the perturbed dynamics deviates from the unperturbed behavior according to~\eqref{7}.

A first major insight, discussed in detail in Sec.~\ref{s4} and derived in Appendix~\ref{app3},
is that $g(t)$
solves the non-linear integro-differential equation
\begin{equation}
\label{eq:gIntEq}
	\dot{g}(t) = -\lambda^2 \int_0^t ds \; g(t - s) \, g(s) \, \tilde\sigma^2(s)
\end{equation}
with the initial condition $g(0) = 1$,
where $\dot g(t)$ is the derivative of $g(t)$ and
\begin{equation}
\label{eq:pertProfFT}
	\tilde\sigma^2(t) := \int \frac{d E}{\varepsilon} \, e^{i E t} \, \sigma^2(E)
\end{equation}
is the Fourier transform of the perturbation profile from~\eqref{5}.
Notably, this result resembles common relations for response functions,
but is distinctly non-linear.
Compared to the indirect representation in terms of Eqs.~\eqref{8} and~\eqref{9},
the new Eq.~\eqref{eq:gIntEq} exposes the relationship between the system's response to the perturbation and the perturbation profile $\sigma^2(x)$ much more clearly.
Moreover, Eq.~\eqref{eq:gIntEq} considerably simplifies the quantitative evaluation of $g(t)$ for general $\sigma^2(x)$ since it can be solved numerically by standard integration techniques.
Not least, it will turn out useful in Sec.~\ref{s42} to assess the asymptotic behavior of $g(t)$ for small and large $t$.

Our second major insight is an analytical approximation for $g(t)$ that covers practically all cases of physical relevance and depends on just two parameters of the perturbation $V$.
The first one quantifies its \emph{intrinsic strength}
\begin{eqnarray}
\alpha:=\sigma^2(0)/\lvsp
\label{73}
\end{eqnarray}
by relating the magnitude $\sigma^2(0)$ of the matrix elements $V^0_{mn}$ for close-by energy levels [cf.\ Eq.~\eqref{5}] to the mean level spacing $\varepsilon$ of the unperturbed Hamiltonian [cf.\ above Eq.~\eqref{4}].
In view of Eq.~\eqref{1}, the \emph{overall perturbation strength} is thus characterized by $\alpha \lambda^2$.
The second parameter assesses the perturbation's energy range or \emph{band width}
\begin{eqnarray}
\Delta_v :=\frac{1}{\sigma^2(0)}\int_0^\infty dE\, \sigma^2(E)
\ .
\label{72}
\end{eqnarray}
Note that we will also admit cases where the matrix is not 
banded at all, or where the perturbation
profile $\sigma^2(E)$ decays 
only very slowly with $E$, so that $\Delta_v$ in (\ref{72})
is infinitely large.

From these two parameters, we can derive the ``golden-rule rate'' [see the remarks below Eq.~\eqref{80} for an explanation of the name]
\begin{eqnarray}
\Gamma := 2\pi\lambda^2\alpha
\label{77}
\end{eqnarray}
as well as the three auxiliary rates
\begin{equation}
\label{82}
\gamma_n :=
		\frac{2 \Delta_v}{\pi} \left[ 1 \pm n \sqrt{1 -  \frac{\pi\Gamma}{2 \Delta_v}}  \right]
\end{equation}
for $n = -1, 0, 1$.
With these definitions, our main result is that $g(t)$ can be excellently approximated for very general perturbation profiles $\sigma^2(x)$ by
\begin{equation}
\label{81}
	g(t) = \frac{ (\gamma_+ - \frac{\Gamma}{2} ) e^{ -\gamma_- \lvert t \rvert } 
	- \Gamma e^{-\gamma_0 \lvert t \rvert} + (\gamma_{-} - \frac{\Gamma}{2} ) 
	e^{-\gamma_+ \lvert t \rvert } }{ 2 (\gamma_0 - \Gamma) } \,,
\end{equation}
where $\gamma_+ \equiv \gamma_1$ and $\gamma_- \equiv \gamma_{-1}$ is understood.
This relation will be derived first in Sec.~\ref{s3} as a suitable truncation of the exact result for a special class of perturbation profiles $\sigma^2(x)$.
The generalization to largely arbitrary $\sigma^2(x)$ will be achieved in Sec.~\ref{s4} by establishing that $g(t)$ is rather insensitive to further details of $\sigma^2(x)$ beyond the parameters $\alpha$ from~\eqref{73} and $\Delta_v$ from~\eqref{72}.
Finally, the validity of Eq.~\eqref{7} in combination with the approximation~\eqref{81} will also be demonstrated in concrete example systems in Sec.~\ref{s6}.

%%%%%%%%%%%%%%%%%%%%%%%%%%%%%%%%%%%%%%%%%%%%%%%%%%%%
%%%%%%%%%%%%%%%%%%%%%%%%%%%%%%%%%%%%%%%%%%%%%%%%%%%%
\section{Analytical approximations}
\label{s3}
For later convenience, we introduce the abbreviation
\begin{eqnarray}
f(x):=\frac{\lambda^2}{\varepsilon}\sigma^2(x)
\ ,
\label{10}
\end{eqnarray}
and -- in view of its physical meaning as discussed around (\ref{5}) --
we will often employ the name {\em perturbation profile} not only for $\sigma^2(x)$, but also for this function $f(x)$.
The integral equation (\ref{9}) thus takes the form
\begin{eqnarray}
G(z)\,\left[z-\int dx\, G(z-x)f(x)\right] = 1 
\ .
\label{11}
\end{eqnarray}
Our goal in this
section is to find
(approximate) solutions of (\ref{9}) by analytical 
means for the special class of functions $f(x)$
which can be written in the form
\begin{eqnarray}
f(x)=\sum_{n=1}^N \frac{f_n}{1+(x/a_n)^2}
\label{12}
\end{eqnarray}
with $f_n\in\RR$ and pairwise different $a_n\in\RR^+$.
The implications of these findings for more general
perturbation profiles $f(x)$ will be discussed in Sec.~\ref{s4}.

Before going into the actual calculations, we outline 
the underlying general strategy of our approach:
Introducing the complex half-plane
\begin{eqnarray}
\CC^-:=\{ z\in\CC\, | \, \mbox{Im}(z)<0 \}
\ ,
\label{13}
\end{eqnarray}
our starting point is the assumption that $G(z)$
exhibits the following two properties (at least)
for all $z\in \CC^-$:
\\[0.3cm]
(i) $G(z)$ is analytic. In particular, $G(z)$ thus
exists (is well-defined) and is finite (exhibits no singularities).
\\[0.3cm]
(ii) $|G(z)|\to 0$ for $|z|\to\infty$.
\\[0.3cm]
Under these assumptions, and given some 
perturbation profile
$f(x)$ of the form (\ref{12}), we will then 
determine approximative solutions of (\ref{11}),
at least for all $z\in\CC^-$, and verify that they
indeed fulfill the initial assumptions (i) and (ii).
Moreover, these approximations can be systematically 
improved
and converge to a (formally) exact solution,
which again fulfills (i) and (ii). 
Accordingly, self-consistent approximate as 
well as exact solutions will be obtained. 
Finally, we will also evaluate their Fourier transform
$g(t)$ in (\ref{8}).
The question whether these solutions of (\ref{11})
are unique is quite difficult and addressed in somewhat 
more detail in Appendix \ref{app1}.

To begin with, we choose an arbitrary but fixed $\eta>0$ 
and define
\begin{eqnarray}
\CC^+_\eta & := & \{ z\in\CC\, | \, \mbox{Im}(z) > -\eta \}
\ ,
\label{14}
\\
\CC^-_\eta & := & \{ z\in\CC\, | \, \mbox{Im}(z)\leq -\eta \} 
\ .
\label{15}
\end{eqnarray}
Since we will later only need asymptotically small
$\eta$'s in (\ref{8}), we furthermore can and will
assume that $\eta<a_n$ for all $n$.

Focusing on an arbitrary but fixed $y\in\CC^-_\eta$, it
follows that $y-z\in \CC^-$ for any $z\in \CC^+_\eta$.
Hence, $F(z):=G(y-z)\, f(z)$ is a well-defined analytic function 
on $\CC^+_\eta$ up to simple poles at $z=i a_n$,
which furthermore satisfies $|z^2F(z)|\to 0$ for all 
$z\in\CC^+_\eta$ when $|z|\to\infty$.
Under these premises, the integral $\int \!dx \, F(x)$ can be readily evaluated by 
textbook residue techniques, yielding for any given 
$y\in\CC^-_\eta$ the result
\begin{eqnarray}
\int dx\, G(y-x) f(x) & = & \sum_{n=1}^N \tilde f_n a_n G(y-ia_n)
\ ,
\label{16}
\\
\tilde f_n & := & \pi\, f_n
\ .
\label{17}
\end{eqnarray}
Finally, the integral equation (\ref{11}) can thus be rewritten
in the form
\begin{eqnarray}
G(z) = \frac{1}{z-\sum_{n=1}^N \tilde f_n a_n G(z-ia_n)}
\label{18}
\end{eqnarray}
for all $z\in\CC^-_\eta$. Since $\eta$ in (\ref{15}) may be
arbitrarily small, even arbitrary $z\in\CC^-$ are actually
admitted in (\ref{18}).

The remaining task will be to show that a well-defined solution
$G(z)$ of (\ref{18}) exists for all $z\in\CC^-$ and
that it satisfies the above requirements (i) and (ii).
A second key objective will be to determine
approximate solutions of this equation (\ref{18}).
The detailed procedure will first be illustrated
in the simplest case with $N=1$ in (\ref{12}), 
while for $N>1$ the generalization of
the main results will be provided without
repeating all details.

%%%%%%%%%%%%%%%%%%%%%%%%%%%%%%%%%%%%%%%%%%%%%%%%%%%%%%%
\subsection{Special case $N=1$}
\label{s31}

Focusing on $N=1$ in (\ref{12}), and adopting the
abbreviations 
\begin{eqnarray}
a & := & a_1
\ ,
\label{19}
\\
\beta & := & \tilde f_1/a
\ ,
\label{20}
\\ 
H(z) & := & -ia\, G(-iaz)
\ ,
\label{21}
\end{eqnarray}
we can rewrite (\ref{18}) as
\begin{eqnarray}
H(z) = \frac{1}{z+ \beta\, H(z+1)}
\ .
\label{22}
\end{eqnarray}
In view of Eq.~\eqref{12} and the definitions \eqref{73}, \eqref{72}, and \eqref{10},
the parameter $\beta$ can be written as $\beta = \pi^2 \alpha \lambda^2 / 2 \Delta_v$ and thus relates the overall perturbation strength $\alpha \lambda^2$ [see below Eq.~\eqref{73}] to the band width $\Delta_v$.
Hence $\beta \ll 1$ corresponds to weak perturbations, whereas $\beta \gg 1$ amounts to strong perturbations (see also Sec.~\ref{s4}).

Analogously to the paragraph below (\ref{18}),
we are now looking for (approximate) solutions of 
(\ref{22}) which are, at least for all $z\in\CC$ with 
$\mbox{Re}(z)>0$, well-defined (exist), analytic, and satisfy 
$|H(z)|\to 0$ for $|z|\to\infty$.
We also remark that,
due to (\ref{5}), (\ref{10}), and (\ref{12}),
we can and will restrict ourselves to
real and positive values of $\beta$ in (\ref{20}).

Upon iteration of Eq. (\ref{22}) one readily obtains
\begin{eqnarray}
H(z) = \frac{1}{z+ \beta\, \frac{1}{z+1+\beta\,\frac{1}{z+2+\beta \cdots}}}
\ ,
\label{23}
\end{eqnarray}
where it is {\em a priori} understood that the iteration ends 
after $k$ steps with a term $\beta\, H(z+k)$ in the last 
denominator.
However, for any given $z$ with $\mbox{Re}(z)>0$
and any $\beta\in\RR^+$ it is rigorously shown in Appendix~\ref{app2}
that the right-hand side of~\eqref{23} converges towards a 
well-defined limit as the number of steps 
tends to infinity.
In this sense, the function $H(z)$ in (\ref{23})
exists (is well-defined) as an infinite continued 
fraction and is a solution of the original Eq.~(\ref{22}).
As already mentioned above Eq. (\ref{14}), 
its uniqueness is here tacitly taken for granted 
and discussed in somewhat more detail in 
Appendix \ref{app1}.
In turn, this yields a formally exact solution 
$G(z)$ of~\eqref{11} via~\eqref{21}.
Finally, we can conclude that by truncating those
continued fraction solutions after a finite number 
of steps, one obtains approximations which can be 
systematically improved by including more steps
(see also the numerical examples in Fig.~\ref{fig:G} below).

In view of the above existence proof of 
$H(z)$ in (\ref{23}), it is furthermore quite reasonable 
to expect that also with respect 
to other properties of $H(z)$ nothing ``dangerous'' will 
happen in any of the nested denominators of the continued 
fraction expression, at least for any $z$ with $\mbox{Re}(z)>0$
and $\beta\in\RR^+$.
Specifically, it is quite plausible that $H(z)$ will be analytic for
any such $z$. Likewise, upon increasing $|z|$, each dominator 
grows (in modulus), hence one expects that $H(z)$ 
approaches zero.
In other words, the requirements below (\ref{22}) are satisfied.

From (\ref{23}) it follows that $H^\ast(z)=H(z^\ast)$, 
implying for $G(z)$ according to (\ref{21}) the 
``symmetry property''
\begin{eqnarray}
G^\ast(z)=-G(-z^\ast)
\ .
\label{24}
\end{eqnarray}
For the usual decomposition of $z$ and $G(z)$ 
into real and imaginary parts,
\begin{eqnarray}
G(x+iy) = v(x,y)+i w(x,y)
\ ,
\label{25}
\end{eqnarray}
the corresponding symmetries of the real and imaginary 
parts of $G(z)$ thus take the form
\begin{eqnarray}
v(-x,y) & = & -v(x,y)
\ ,
\label{26}
\\
w(-x,y) & = & w(x,y)
\ .
\label{27}
\end{eqnarray}
Exploiting those symmetries in (\ref{8}), 
one readily can conclude, as detailed in 
Appendix \ref{app3},
that
\begin{eqnarray}
g(t) & = & \frac{1}{2\pi i}\,  \lim_{\eta\downarrow0}
\int dx\, G(x-i\eta) \, e^{ix|t|}
\label{28}
\end{eqnarray}
and that $g(t)$ is an even and real-valued function of $t$.

Side remark: Later the integral in (\ref{28})
will be evaluated by residue methods.
Since $G(z)$ is so far only assumed to be analytic for
$z\in\CC^-$, a finite $\eta>0$ under the integral 
is then still needed in principle.
In practice, we will actually evaluate the integral mostly for
certain approximations of $G(z)$ which will be everywhere
analytic up to isolated poles with strictly positive imaginary parts.
In this case the limit $\eta\to 0$ can be performed before the
integration.
However, in the large $\beta$ limit (see Sec. \ref{s32})
we will also encounter an 
example where $G(z)$ is indeed non-analytic on almost the entire 
real axis. In such a case, keeping $\eta$ finite under the intergral
is indispensable. 
Somewhat related issues are also addressed in Appendix \ref{app1}.

Finally, at least for sufficiently small $\beta$, a very natural sequence of
better and better approximations arises by truncating the continued fraction
in (\ref{23}) later and later. 
In particular, the first order approximation takes the form
\begin{eqnarray}
H(z) = \frac{1}{z+ \beta\, \frac{1}{z+1}}
\label{29}
\end{eqnarray}
and likewise for the second order approximation
\begin{eqnarray}
H(z) = \frac{1}{z+ \beta\, \frac{1}{z+1+\beta\,\frac{1}{z+2}}}
\ ,
\label{30}
\end{eqnarray}
and so on.
The corresponding ``zeroth order'' approximation $H(z)=1/z$ turns 
out to be of little use. On the other hand, especially for small $|z|$
also other kinds of approximations, such as $1/[z+\beta]$ or
$1/[z+\beta/(z+1+\beta/2)]$ etc. could be considered.
Generally, we found that the latter approximations are essentially
of the same quality as those in (\ref{29}), (\ref{30}), and hence we do not 
further pursue them here.

Next, we rewrite the first order approximation (\ref{29}) as
\begin{eqnarray}
H(z) & = & \frac{z+1}{(z+x_1)(z+x_2)}
\ ,
\label{31}
\\
x_{1,2} & := & \frac{1\pm\sqrt{1-4\beta}}{2}
\ .
\label{32}
\end{eqnarray}
This approximative solution is thus analytic
(at least) for all $z\in\CC$ with $\mbox{Re}(z)>0$ 
(since $\mbox{Re}(x_{1,2})>0$ for any $\beta>0$), 
and approaches zero for large $|z|$,
i.e., it still satisfies the requirements below (\ref{22}).

The corresponding approximation for $G(z)$ is recovered upon 
introducing (\ref{31}) into (\ref{21}), yielding
\begin{eqnarray}
G(z)=\frac{z-ia}{(z-i a x_1)(z-i a x_2)}
\ .
\label{33}
\end{eqnarray}
Finally, inserting (\ref{33}) into (\ref{28}) yields by means of
standard residue techniques the result
\begin{eqnarray}
g(t) 
& = &
\frac{x_1\, e^{-ax_2|t|} - x_2\, e^{-ax_1|t|}}{x_1-x_2}
\ .
\label{34}
\end{eqnarray}

While the differentiablilty properties of  
(\ref{34}) are obvious for all $t\not=0$, 
the time-point $t=0$ warrants a closer look.
Exploiting (\ref{32}),
one readily confirms that
\begin{eqnarray}
g(0) & = & 1\ ,
\label{35}
\\
\dot g(0)  & = & 0\ ,
\label{36}
\\
\ddot g(0)  & = & - a^2\beta \ ,
\label{36a}
\end{eqnarray}
and that the third derivative of $g(t)$ does 
not exist at $t=0$ (the second derivative is continuous but not differentiable).

Since $\mbox{Re}(x_{1,2})>0$ for any $\beta>0$
in (\ref{32}), the right hand side of (\ref{34}) approaches 
zero for large $t$.
In the case $\beta<1/4$, Eq. (\ref{32}) implies $x_1>x_2>0$,
hence the large-$t$ asymptotics is dominated by the
first summand on the right hand side of (\ref{34}).
For $\beta > 1/4$, the two roots in (\ref{32}) turn 
complex, hence (\ref{34}) amounts to an exponential 
decay $e^{-a|t|/2}$ times
a sinusoidal oscillation.
Moreover, for small $\beta$ one readily recovers
the asymptotic approximation
\begin{eqnarray}
g(t) & \simeq & e^{-\Gamma |t|/2}
\ ,
\label{37}
\end{eqnarray}
where $\Gamma = 2\pi \lambda^2 \alpha = 2\pi f_1$ according to~\eqref{73}, \eqref{77}, (\ref{10}) and~\eqref{12}.
Observing that this approximation
violates (\ref{36}), it follows that 
the limits $t\to 0$ and $\beta\to\ 0$ do not commute.

Turning to the second order approximation from 
(\ref{30}), one can proceed essentially like before.
One important observation is that, incidentally,
$H(z)$ happens to exhibits a simple pole at
$z=-1$,
as can be immediately verified by closer
inspection of (\ref{30}).
The remaining two poles then readily
follow as the solutions of a quadratic equation.
Along these lines, one finally arrives at
the second order approximation
\begin{eqnarray}
g(t) 
\!\!\!
&=&
\!\!\!
\frac{
(x_1-\beta)  e^{-ax_2|t|}
+
(x_2-\beta)  e^{-ax_1|t|}
-
2\beta  e^{-a|t|}
}{2(1-2\beta)}
\qquad
\label{39}
\\
x_{1,2}
\!\!
&:=&
\!\!
1\pm\sqrt{1-2\beta}
\ ,
\label{40}
\end{eqnarray}
where $\beta$ 
is
again defined by (\ref{20}). 
Likewise, one readily recovers once again 
the properties in and around Eqs. (\ref{35})-(\ref{37}).

%%%%%%%%%%%%%%%%%%%%%%%%%%%%%%%%%%%%%%%%%%%%%%%%%%%%%%
\begin{figure}
\includegraphics[scale=1]{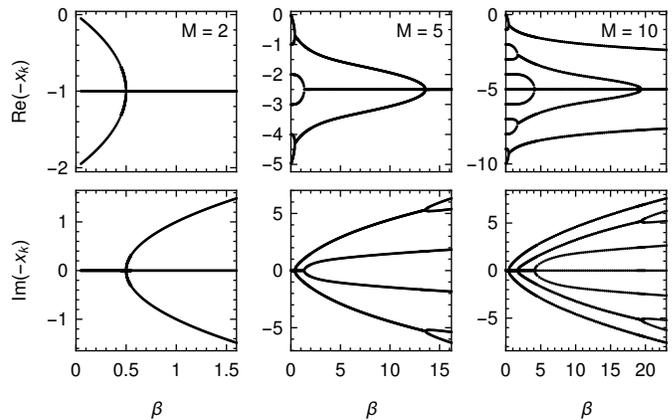}
\caption{
Real and imaginary parts of the $M+1$ poles versus $\beta$
for the order-$M$ approximation of $H(z)$ from (\ref{23}).
}
\label{fig:HPoles}
\end{figure}

%%%%%%%%%%%%%%%%%%%%%%%%%%%%%%%%%%%%%%%%%%%%%%%%%%%%%%

%%%%%%%%%%%%%%%%%%%%%%%%%%%%%%%%%%%%%%%%%%%%%%%%%%%%%%
\begin{figure*}
\includegraphics[scale=1]{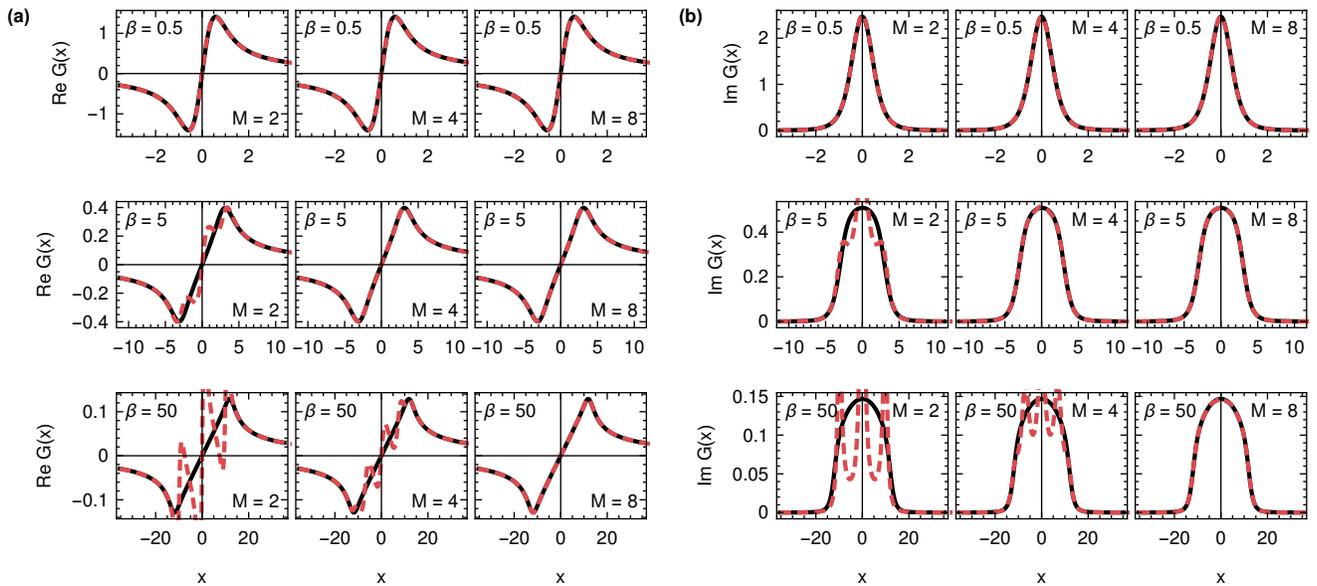}
\caption{
(a) Real and (b) imaginary parts of the order-$M$ approximations
of $G(z)$ for real-valued arguments $z=x$ (dashed, red) together with the
corresponding numerically obtained exact solutions (solid, black) of the 
integral equation (\ref{11}) with $f(x)$ from~\eqref{12}, $N = 1$, and 
$a=a_1=1$.
}
\label{fig:G}
\end{figure*}
%%%%%%%%%%%%%%%%%%%%%%%%%%%%%%%%%%%%%%%%%%%%%%%%%%%%%%

We conclude with some remarks regarding the general
structure of the higher-order approximations of the continued-fraction expression (\ref{30}).
Adopting the letter $M$ for the order of the approximation,
the first order approximation in (\ref{29}) thus corresponds to
$M=1$, the second order approximation in (\ref{30}) to $M=2$
and so on.
Starting with $M=1$, we found that for small-to-moderate 
$\beta$, both ``roots'' $x_{1,2}$ in (\ref{32}) are 
real and positive, one being close to zero and one 
close to unity. 
At $\beta=1/4$ they turn into a complex conjugated pair
with real part $1/2$.
The same properties apply to the corresponding 
poles at $-x_{1,2}$ of $H(z)$ from (\ref{31}).
Analogously, for $M=2$ the two poles at $-x_{1,2}$
are, for small $\beta$, real and negative 
according to (\ref{40}), one being close to 
zero and one close to $-2$.
As $\beta$ exceeds the value $1/2$, they again turn into
a complex conjugated pair with real 
part $-1$. 
As discussed above (\ref{39}), there is a third
pole which happens to be exactly at $-1$ for all 
$\beta$.
More generally, it thus seems reasonable to 
expect that the order-$M$ approximation of (\ref{23}) 
will exhibit $M+1$ real- and positive-valued
poles close to $0,-1,-2,...,-M$ for small $\beta$,
which turn step by step into complex conjugated 
pairs upon increasing $\beta$.
While a more detailed analytical elaboration of this 
issue seems quite cumbersome, its numerical 
exploration is straightforward.
Fig.~\ref{fig:HPoles} depicts the so-obtained numerical
findings for several $M$ values, essentially
confirming our above expectations.
Moreover, for any given $M$ the real part of all poles 
apparently assumes the same value $-M/2$ for 
sufficiently large $\beta$.

Likewise, with respect to the corresponding order-$M$ approximations for $G(z)$ according to (\ref{21}),
the findings in Fig.~\ref{fig:G} quite convincingly
demonstrate convergence of the continued-fraction
representation for arbitrary $\beta$ 
towards the solutions of the original integral 
equation (\ref{11}), which we numerically determined
as explained in more detail in Ref.~\cite{dab20a}.
In particular, one expects that all those approximations
for $G(z)$ will again satisfy the two requirements 
below (\ref{13}) and the symmetry (\ref{24}), 
and that the corresponding Fourier transform $g(t)$ 
from (\ref{28}) will be a sum of $M+1$ exponentially 
decaying functions with the properties (\ref{35})-(\ref{37}) 
and with decay rates which
are given by the $M+1$ poles from above.

Taking for granted that $G(z)$ satisfies the
two requirements below (\ref{13}) even on the entire 
lower complex half plane (which is reasonable to expect
from the above discussion of the poles of $G(z)$), 
it follows that the real and imaginary parts of $G(z)$
are connected via Kramers-Kronig relations, in agreement
with how they are seen to behave in Fig.~\ref{fig:G}.

\begin{figure*}
\includegraphics[scale=1]{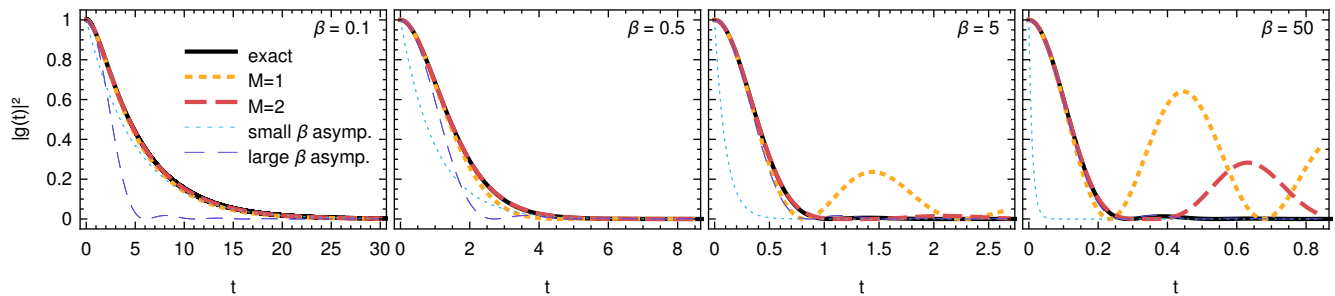}
\caption{
Analytic approximations of $\lvert g(t) \rvert^2$ 
versus numerically exact results
for the perturbation profile~\eqref{12} with $N = 1$, 
$a=a_1=1$, and various values of $\beta$
(note the different scales of the $t$-axis).
Bold solid lines: 
Numerically exact solutions of (\ref{8}) and (\ref{9}).
Bold dotted lines: First order approximations from (\ref{34}).
Bold dashed lines: Second order approximations from (\ref{39}).
Thin dotted lines: Small $\beta$ asymptotics from (\ref{37}).
Thin dashed lines: Large $\beta$ asymptotics from (\ref{48}).
}
\label{fig:gLorentz}
\end{figure*}

Finally, the first- and second-order approximations 
from (\ref{34}) and (\ref{39}) are compared with
the numerically exact behavior in Fig.~\ref{fig:gLorentz}.
More precisely speaking, $|g(t)|^2$ rather 
than $g(t)$ itself is plotted since it is this 
quantity which actually matters in (\ref{7}).
Clearly, the agreement is practically perfect for the 
second-order approximation, apart from very large 
$\beta$-values and times considerably larger than 
the relaxation time of the exact $|g(t)|^2$.
The latter shortcoming could be 
readily remedied by switching
to the large-$\beta$ approximation (thin dashed lines, 
see Eq.~\eqref{48} below)
when $a^2\beta$ exceeds some critical 
value, e.g., $a^2\beta>10$.
Moreover, also the higher-order approximations would
exhibit even much smaller such deviations,
hence they are not shown.

The very good performance of the second order
approximation for $g(t)$ is remarkable for two reasons.
First, in its original derivation we assumed that
$\beta$ is small. Second, the corresponding $M = 2$ 
approximations for $G(x)$ in Fig.~\ref{fig:G} indeed
show much more pronounced deviations for larger values of $\beta$.
We observe that these deviations are most striking in the region 
around $x = 0$, whereas the tails are generally reproduced well 
already for small $M$ and large $\beta$.
Since $g(t)$ is essentially the Fourier transform of 
$\operatorname{Im} G(x)$ [see Eq.~\eqref{28}], the short-time 
behavior is mainly determined by those tails and the deviations 
for small $x$ become effective only later when 
the exact $\lvert g(t) \rvert^2$ 
has basically relaxed to zero for all practical purposes
(cf.\ Fig.~\ref{fig:gLorentz}).

%%%%%%%%%%%%%%%%%%%%%%%%%%%%%%%%%%%%%%%%%%%%%%%%%%%%%%

%%%%%%%%%%%%%%%%%%%%%%%%%%%%%%%%%%%%%%%%%%%%%%%%%%%%%%

%%%%%%%%%%%%%%%%%%%%%%%%%%%%%%%%%%%%%%%%%%%%%%%%%%%%%%%
\subsection{Large $\beta$ asymptotics for $N=1$}
\label{s32}

With the definition
\begin{eqnarray}
I(z):=\sqrt{\beta}\, H(\sqrt{\beta} z)
\label{41}
\end{eqnarray}
we can rewrite (\ref{22}) as
\begin{eqnarray}
I(z)=\frac{1}{z+I(z+1/\sqrt{\beta})}
\ .
\label{42}
\end{eqnarray}
For asymptotically large $\beta$ we thus can conclude that
\begin{eqnarray}
I(z)[I(z)+z]-1=0
\label{43}
\end{eqnarray}
and therefore
\begin{eqnarray}
I(z)=\frac{-z\pm\sqrt{4+z^2}}{2}
\ .
\label{44}
\end{eqnarray}
Returning to $G(z)$ via (\ref{41}) and (\ref{21})
yields
\begin{eqnarray}
G(z) & = & \frac{2}{\gamma^2}\left[ z\pm i\sqrt{\gamma^2-z^2}\right]
\ ,
\label{45}
\\
\gamma & := & 2a\sqrt{\beta}
 \ .
\label{46}
\end{eqnarray}
As usual, the complex square root is defined 
here as $\sqrt{z}:=\exp\{\ln(z)/2\}$, 
where $\ln(z):=\ln(|z|)+i\arg(z)$, and where $\arg(z)$ 
is the principal value argument of the complex 
number $z$.
Hence, the right hand side of~(\ref{45}) is,
for either choice of the sign, 
analytic apart from cuts on 
$(1,\infty)$ and $(-\infty,-1)$,
i.e., the second requirement below 
(\ref{13}) is fulfilled.
However, the first requirement below (\ref{13}),
$|G(z)|\to 0$ for $z\in\CC^-$ with $|z|\to\infty$, 
is only fulfilled when choosing the 
plus sign in (\ref{45}), resulting in
\begin{eqnarray}
G(z) & = & \frac{2}{\gamma^2}\left[ z + i\sqrt{\gamma^2-z^2}\right]
\ .
\label{47}
\end{eqnarray}
Finally, also the symmetry property (\ref{24})
is again fulfilled, as can be inferred from
(\ref{47}) by exploiting that $\sqrt{z^\ast}=(\sqrt{z})^\ast$.

The corresponding function $g(t)$ is most
conveniently obtained by introducing (\ref{47})
directly into the definition (\ref{8}), yielding
\begin{equation}
g(t) = 2 J_1(\gamma t)/\gamma t
\ ,
\label{48}
\end{equation}
where $J_1(x)$ is the Bessel function of the first kind of order $1$.
Remarkably, one thus recovers once again exactly the same 
properties as in (\ref{35})-(\ref{36a}).
But in contrast to the remark below (\ref{36a}), 
now also all higher derivatives of $g(t)$ exist at $t=0$.
Finally, the large-$t$ asymptotics of (\ref{48})
amounts to a decay proportional to
$|t|^{-3/2}$ times a sinusoidal oscillation, again 
somewhat similar to the findings for large 
$\beta$ above (\ref{37}),
albeit the decay is now given by a power law rather than an exponential.
In fact,
as will be demonstrated in more detail in Sec.~\ref{s42} [see the discussion below Eq.~\eqref{eq:gLargeTimeRate}],
the limits $\beta \to \infty$ and $t \to \infty$ do not commute,
so the present $\lvert t \rvert^{-3/2}$ asymptotics will ultimately cross over to an exponential decay
for any preset (finite) value of $\beta$.

One possibility to determine the first order correction
to the above asymptotics (\ref{47}) 
would be to rewrite (\ref{42}) as
\begin{eqnarray}
& & I(z)[I(z) + d(z) +z]-1=0
\ ,
\label{49}
\\
& & d(z) := I(z+1/\sqrt{\beta})-I(z)
\ ,
\label{50}
\end{eqnarray}
and then to approximate the small difference on the right
hand side of (\ref{50}) by means of the previous 
approximation from (\ref{47}).
But since the integral in (\ref{8}) is then likely to only 
be tractable by numerical means, we do not further 
pursue this issue.

%%%%%%%%%%%%%%%%%%%%%%%%%%%%%%%%%%%%%%%%%%%%%%%%%%%%%%%
\subsection{General $N$}
\label{s33}

Next we briefly sketch the generalizations of the
so far employed considerations when going over from $N=1$ 
to arbitrary $N$ in (\ref{12}):
In terms of the same auxiliary function $H(z)$ as in
(\ref{21}) one readily obtains from (\ref{18}) as a generalization of
(\ref{23}) the continued fraction form
\begin{eqnarray}
H(z) \!\! & = & \!\! 
\frac{1}{z + \sum_{n=1}^N \beta_n H(z+\tilde a_n)}
\label{51}
\\
& = & \!\! \frac{1}{z + \sum_{n=1}^N 
\beta_n \frac{1}{z+\tilde a_n + \sum_{m=1}^N
\beta_m \frac{1}{z+ \tilde a_n+\tilde a_m + \cdots}}}
\ \ \ \ \
\label{52}
\\[0.1cm]
\tilde a_n \!\! & := & \!\! a_n/a
\ ,
\ \
\label{53}
\\[0.1cm]
\beta_n & := & \tilde f_n a_n/a^2
\ .
\label{54}
\end{eqnarray}
Here, the quantity $a$ can still be chosen arbitrarily.
For instance, the previously considered special case with
$N=1$ is readily recovered by choosing $a=a_1$, implying
$\tilde a_1=1$ and $\beta_1=\tilde f_1/a$ (cf.~(\ref{20})).
Another natural choice is $a=1$, implying $\tilde a_n=a_n$
and $\beta_n=\tilde f_n a_n$.
For the rest, it is reasonable to expected that conclusions
analogous to those below (\ref{23}) will also apply 
to the function $H(z)$ in (\ref{52}).
Most importantly, Eqs. (\ref{52}) and
(\ref{21}) again imply the same symmetry property 
of $G(z)$ as in (\ref{24}) and hence the same relation
for $g(t)$ as in (\ref{28}).

Focusing on sufficiently small $\beta_n$,
the generalization of the previous first-order 
approximation (\ref{29}) now takes the form
\begin{eqnarray}
H(z) & = & \frac{1}{z+ \sum_{n=1}^N 
\beta_n \frac{1}{z+\tilde a_n}} 
=\frac{1}{-h(-z)}
\label{55}
\\
h(z) & := & z+\sum_{n=1}^N\frac{\beta_n}{z - \tilde a_n} 
\label{56}
\end{eqnarray}

By means of a graphical sketch of the function 
$h(x)$ for $x\in\RR$, one readily
sees that the equation $h(x)=0$ generically exhibits $N+1$
solutions, which are all real-valued and positive, and 
which we denote by $x_k$ with $k=1,...,N+1$,
i.e.,
\begin{eqnarray}
h(x_k)=0 \ \mbox{with $x_k\in\RR^+$ and $k=1,...,N+1$.}
\label{57}
\end{eqnarray}
Furthermore, for sufficiently small $\beta_n$,
one finds that one solution of $h(x)=0$
is close to zero, and that all further 
solutions are close to one of the 
$\tilde a_n$'s.
Without loss of generality, we may thus choose
the labels so that $x_{N+1}\simeq 0$ and
$x_n\simeq \tilde a_n$ for $n=1,...,N$.
Accordingly, it is natural to make the ansatz
$x_n=-\tilde a_n+\delta_n+\delta_n'+...$
and $x_{N+1}=\delta_{N+1}+\delta_{N+1}'+...$,
where the $\delta_k$ (for $k=1,...,N+1$) 
are of first order in the small parameters $\beta_n$, 
the $\delta_k'$ of second order, and the terms ``$+...$'' 
of higher order.
Introducing this ansatz into (\ref{56}), (\ref{57}) 
and solving the equations order by 
order then yields the result 
(up to first order in the $\beta_n$)
\begin{eqnarray}
x_n \simeq \tilde a_n-\frac{\beta_n}{\tilde a_n}\ \ \mbox{for $n=1,...,N$}
\ ,
\label{58}
\end{eqnarray}
and (up to second order in the $\beta_n$)
\begin{eqnarray}
x_{N+1} & \simeq & 
\left(\sum_{n=1}^N \frac{\beta_n}{\tilde a_n}\right)\,\left(1+\sum_{n=1}^N \frac{\beta_n}{\tilde a^2_n}\right)
\ .
\label{59}
\end{eqnarray}

Rewriting (\ref{55}) as
\begin{eqnarray}
H(z) & = & q(z)/p(z)
\ ,
\label{60}
\\
q(z) & := &  \prod_{n=1}^N (z+\tilde a_n)
\ ,
\label{61}
\\
p(z) & := & z\, q(z)+\sum_{n=1}^N \beta_n\prod_{m=1, m\not=n}^N (z+\tilde a_m)
\ ,
\label{62}
\end{eqnarray}
one readily concludes that $p(z)$ 
is a polynomial of the form $z^{N+1}+...$,
whose zeroes must coincide with
the $-x_k$'s from above, i.e.,
\begin{eqnarray}
p(z)=\prod_{k=1}^{N+1} (z+x_k)
\label{63}
\end{eqnarray}
Together with (\ref{21}), (\ref{60}), and (\ref{61}) we thus arrive at
the approximation
\begin{eqnarray}
G(z)=
\frac{\prod_{n=1}^N(z-i \tilde a_n)}{\prod_{k=1}^{N+1} (z-i a x_k)}
\ .
\label{64}
\end{eqnarray}
Finally, the integral in (\ref{28}) can again be evaluated by 
standard residue methods, resulting in
\begin{eqnarray}
g(t)& = &\sum_{k=1}^{N+1} A_k\ e^{-ax_k|t|}
\ ,
\label{65}
\\
A_k & := & \frac{\prod_{n=1}^N(\tilde a_n - x_k)}{\prod_{j=1,j\not=k}^{N+1} (x_j - x_k)}
\ .
\label{66}
\end{eqnarray}

Exploiting (\ref{58}) and (\ref{59}) yields for the amplitudes
$A_k$ in (\ref{66}) the approximations
\begin{eqnarray}
A_n & \simeq & - \frac{\beta_n}{\tilde a_n^2} \ \ \mbox{for $n=1,...,N$}
\ ,
\label{67}
\\
A_{N+1} & \simeq & 1+\sum_{n=1}^N \frac{\beta_n}{\tilde a_n^2}
\ .
\label{68}
\end{eqnarray}
By means of those approximations, one again
recovers 
the properties in and around Eqs. (\ref{35})-(\ref{37})
[see also below Eq. (\ref{40})],
except that $\Gamma$ and $\beta$
now take the form
\begin{eqnarray}
\Gamma & = & 2\pi\,\lambda^2 \frac{\sigma^2(0)}{\varepsilon} = 2\pi \sum_{n=1}^N f_n
\ ,
\label{69}
\\
\beta & := & \sum_{n=1}^N \beta_n
\ .
\label{70}
\end{eqnarray}

Turning finally to the case of asymptotically large $\beta_n$,
we employ the same auxiliary function $I(z)$ as in (\ref{41})
with $\beta$ from (\ref{70}).
Introducing these quantities into (\ref{51})
yields as a generalization of (\ref{42}) the 
result
\begin{eqnarray}
I(z)=\frac{1}{z+\sum_{n=1}^N \frac{\beta_n}{\beta} I(z+\tilde a_n/\sqrt{\beta})}
\ .
\label{71}
\end{eqnarray}
For asymptotically large $\beta$,  all further conclusions
remain exactly the same as in (\ref{43})--(\ref{48}).

%%%%%%%%%%%%%%%%%%%%%%%%%%%%%%%%%%%%%%%%%%%%%%%%%%%%
%%%%%%%%%%%%%%%%%%%%%%%%%%%%%%%%%%%%%%%%%%%%%%%%%%%%
\section{Universality properties of $g(t)$}
\label{s4}

Despite the large variability of admissible perturbations with the properties described in and 
around Eqs.~\eqref{5} and~\eqref{6}, the function $g(t)$, which governs the perturbed 
relaxation in~\eqref{2} and~\eqref{7}, will turn out to be surprisingly universal.
The aim of this section is to track down the origins of this universality.
For this purpose, we will explore how
$g(t)$ varies with the perturbation 
strength and with time.
In both cases, the limiting expressions for small and large values will be found to be 
determined by the overall strength $\alpha \lambda^2$ [see below Eq.~\eqref{73}] and the band width $\Delta_v$ [see Eq.~\eqref{72}]
of the perturbation alone.
In addition, we will inspect several other shapes of the perturbation profile function $\sigma^2(x)$ and the stability of $g(t)$ against those variations.
Exploiting the universality of $g(t)$, we will finally be able to condense the 
perturbation dependence of the prediction~\eqref{7} for the perturbed relaxation 
to just these two parameters,
culminating in the main result~\eqref{81}.

%%%%%%%%%%%%%%%%%%%%%%%%%%%%%%%%%%%%%%%%%%%%%%%%%%%%
\subsection{Asymptotics and crossover with respect to the coupling strength $\lambda$}
\label{s41}

For our special class of functions $f(x)$ from (\ref{12})
one readily finds with (\ref{73}), (\ref{72}), and (\ref{10}) 
the relations
\begin{eqnarray}
\alpha\, \lambda^2& = & \sum_{n=1}^N f_n 
\ ,
\label{74}
\\
\Delta_v & = & \frac{\pi}{2} \frac{\sum_{n=1}^N a_n f_n}{\sum_{n=1}^N f_n}
\ .
\label{75}
\end{eqnarray}
By exploiting (\ref{54}) and (\ref{70}) it follows that
\begin{eqnarray}
\beta & = & 2\lambda^2\alpha\, \Delta_v/a^2
\label{76}
\end{eqnarray}
and with (\ref{36a}) that
\begin{eqnarray}
\ddot g(0)  & = & - 2\lambda^2\alpha\, \Delta_v
\ .
\label{36b}
\end{eqnarray}
Together with (\ref{69}) we thus recover the definition $\Gamma := 2\pi \lambda^2 \alpha$ from~\eqref{77},
and with (\ref{46}) we find that
\begin{eqnarray}
\gamma = \lambda \sqrt{8\alpha\Delta_v} \ .
\label{78}
\end{eqnarray}

For fixed $\alpha$ and $\Delta_v$, it follows from~\eqref{76}
that the regimes of small and large $\beta$ 
are equivalent to small and large coupling strengths 
$\lambda$ in (\ref{1}), respectively [see also the comment below Eq.~\eqref{22}].
Hence (\ref{37}) now applies for small $\lambda$ and is,
according to  (\ref{77}), entirely determined by the parameter
$\alpha$, independently of any further details
of the function $f(x)$ in (\ref{12}).
Likewise, (\ref{48}) applies for large $\lambda$ and 
solely depends on the parameters $\alpha$ and 
$\Delta_v$ via (\ref{78}).

As expected in view of (\ref{37}) and (\ref{48}), 
and elaborated in somewhat more detail in Ref.~\cite{dab20},
the ``crossover'' between the two asymptotic regimes
is roughly determined by the condition 
$\Gamma=\gamma$.
According to (\ref{77}) and (\ref{78}), the corresponding
crossover coupling strength is thus given by 
\begin{eqnarray}
\lambda_c := \sqrt{2\Delta_v/\pi^{2}\alpha}
\ ,
\label{79}
\end{eqnarray}
i.e., the asymptotics from (\ref{37}) and (\ref{48})
are expected to apply for $\lambda\ll\lambda_c$ 
and $\lambda\gg\lambda_c$, respectively.

As shown in more detail in Ref.~\cite{dab20a},
the above discussed asymptotics for large and small 
$\lambda$ in fact apply not only for functions $f(x)$ in (\ref{10}) 
of the form (\ref{12}), but even for largely arbitrary 
perturbation profiles $f(x)$ and thus $\sigma^2(x)$ 
in (\ref{10}), and likewise for the crossover 
condition $\Gamma\approx\gamma$ implying (\ref{79}).

Furthermore, for perturbations which are not banded 
or only very weakly banded, so that the band width 
in~\eqref{72} is infinitely large (see discussion below (\ref{72})), 
the approximation~\eqref{37} is in view of 
(\ref{79}) expected to actually apply for all values 
of the coupling strength $\lambda$ \cite{f1}
(provided they are still compatible 
with the overall restrictions for $\lambda$
discussed in Sec.~\ref{s2}).

The above predicted structural stability of the 
function $g(t)$ for asymptotically small and large $\lambda$
and of the crossover value from (\ref{79}) are also 
illustrated in Fig.~\ref{fig:gBandProfilesCompare} below, 
where we compare numerical solutions of $g(t)$ for various 
perturbation profiles $\sigma^2(x)$ and coupling strengths $\lambda$.

%%%%%%%%%%%%%%%%%%%%%%%%%%%%%%%%%%%%%%%%%%%%%%%%%%%%
\subsection{Asymptotics and crossover with respect to time}
\label{s42}

Interestingly, an analogous crossover between~\eqref{37} 
and~\eqref{48} can also be observed in the time 
domain for a fixed value of $\lambda$.
Our starting point is
the non-linear integro-differential equation~\eqref{eq:gIntEq},
which can be obtained -- as detailed in Appendix \ref{app3} --
via Fourier transformation of~\eqref{11}.
Exploiting~\eqref{10}, it reads
\begin{eqnarray}
\label{eq:gIntEq2}
	\dot{g}(t) & = & - \int_0^t ds \, g(t - s) \, g(s) \, \tilde f(s) \,,
\\
\label{eq:fFT}
	\tilde f(t) & := & \int dx \, e^{i x t} f(x)
\\
g(0) & = & 1 \ ,
\label{35a}
\end{eqnarray}
see also (\ref{35}). From this equation~\eqref{eq:gIntEq2}, 
it is straightforward to deduce a relation 
between the coefficients of a Taylor expansion of $g(t)$ around $t = 0$ 
and the moments
\begin{equation}
\label{eq:PertProfileMoments}
	F_k := \int dx \, x^k \, f(x)
\end{equation}
of the perturbation profile (if they exist, see also below (\ref{36a})).
For our present purpose, an even simpler argument 
is sufficient: For small $t$, the integrand
$g(t - s)g(s)\tilde f(s)$ in (\ref{eq:gIntEq}) can be approximated in
leading order by $[g(0)]^2 \tilde f(0)$, yielding with (\ref{35a})
\begin{equation}
\label{eq:gSmallTime}
	g(t) \simeq 1 - \tfrac{1}{2} \tilde f(0) \, t^2 
		= 1 - \lambda^2 \alpha \, \Delta_v \, t^2 
       \ ,
\end{equation}
where we exploited \eqref{73}, \eqref{72}, \eqref{10}, and \eqref{eq:fFT} 
in the last equality, and where we tacitly assumed that $\tilde f(0)$ 
and hence the band width $\Delta_v$ are finite.
In particular, this is in agreement with our previous findings
(\ref{35}), (\ref{36}), and
(\ref{36b}) for the special class of functions from (\ref{12}).
Remarkably, even the approximation for large $\lambda$
from (\ref{48})
can be shown to exhibit the same asymptotic behavior 
for small $t$ as in (\ref{eq:gSmallTime}).

On the other hand, in the so far excluded case that
$\tilde f(0)$ and hence the band width $\Delta_v$ 
are infinite, one finds that $\tilde f(t)$ 
in (\ref{eq:fFT}) generically develops a $\delta$-peak at $t=0$.
By means of analogous calculations as before
one then recovers from (\ref{eq:gIntEq})
the same small-$t$ asymptotics as in (\ref{37}), i.e., 
$g(t)\simeq 1-\Gamma\, |t|/2$ with $\Gamma$ from (\ref{77}).
Similarly as below (\ref{37}), we thus arrive at the
important conclusion that {\em the limits $t\to 0$ 
and $\Delta_v\to\infty$ do not commute}.

The asymptotic behavior of $g(t)$ for large $t$ can be extracted 
from~\eqref{eq:gIntEq} by means of the ansatz $g(t) = e^{-r t}$.
Substituting into~\eqref{eq:gIntEq}, we find that
\begin{equation}
\label{eq:gLargeTimeRate}
	r = \int_0^t ds \, \tilde f(s) \to \pi f(0) = \pi \alpha \lambda^2
	\quad (t \to \infty) \,.
\end{equation}
For large $t$, we thus conclude that $g(t) \sim e^{-\pi\alpha \lambda^2 t}$ 
coincides with the limiting expression~\eqref{37} for small $\lambda$ or large 
$\Delta_v$ with $\Gamma$ given in~\eqref{77}.
In contrast, the asymptotic behavior~\eqref{48} 
for large $\lambda$ 
exhibits a power-law decay $\sim t^{-3/2}$ as $t \to \infty$.
Similarly as before, we thus find that
the limits $t \to \infty$ and 
$\lambda \to \infty$ 
do not commute either.
But since too large $\lambda$ values are physically 
unrealistic, the actual main conclusion is that the
asymptotics~\eqref{48} must always become
invalid for sufficiently large $t$.

%%%%%%%%%%%%%%%%%%%%%%%%%%
\begin{figure}
\includegraphics[scale=1]{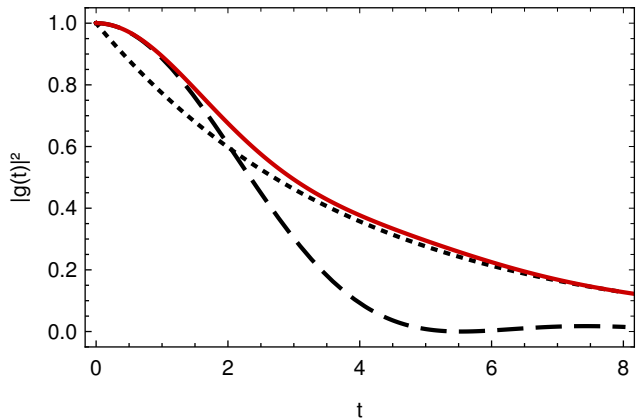}\\
\caption{Numerical solution for $\lvert g(t) \rvert^2$ (solid) 
for the step perturbation profile~$f(x) = 0.04 \, \Theta(1.5^2 - x^2)$ 
with band width $\Delta_v = 1.5$ compared to the 
asymptotic decay characteristics
for weak [dotted, Eq.~\eqref{37}] and strong 
[dashed, Eq.~\eqref{48}] perturbations.
This illustrates the transition in the time domain from the 
Bessel-type behavior for $t \ll t_{\mathrm c}$ to the exponential 
behavior for $t \gg t_{\mathrm c}$ with $t_{\mathrm c} 
\approx \pi / \Delta_v \approx 2.1$.
}
\label{fig:g:tc}
\end{figure}
%%%%%%%%%%%%%%%%%%%%%%%%%%%%%%%%%%%

Similarly as in (\ref{79}),
an estimate for the ``crossover'' time at which the exponential regime sets 
in can be found by equating~\eqref{eq:gSmallTime} with the asymptotic form 
$e^{-\pi \alpha \lambda^2 t} \approx 1 - \pi \alpha \lambda^2 t$, yielding
\begin{eqnarray}
t_{\mathrm c} \approx \pi / \Delta_v
\ ,
\label{80}
\end{eqnarray}
independent of $\lambda$ and $\alpha$ \cite{gen13}.

Altogether, we thus can conclude that the small-$t$ 
asymptotics of $g(t)$ is (for any finite band width $\Delta_v$)
described by~\eqref{48}, while~\eqref{37} captures the
large-$t$ asymptotics, and that the crossover time scale 
is solely determined (to leading order) by the inverse 
band width according to (\ref{80}).
In particular, the two asymptotic approximations in 
(\ref{37}) and (\ref{48}) actually do not only apply 
to small and large $\lambda$, but also to large 
and small $t$, respectively.

At this point it is
noteworthy that Eq.~\eqref{7} entails, incidentally, a non-perturbative justification of Fermi's golden rule \cite{sak94} in the present many-body setting \cite{mal19thermal}.
More precisely, our universal long-time asymptotics with $\lvert g(t) \rvert^2 \sim e^{-\Gamma t}$
and $\Gamma = 2\pi \, \lambda^2 \sigma^2(0) \, \varepsilon^{-1}$ [see Eqs.~\eqref{73} and \eqref{eq:gLargeTimeRate}]
comprises as a special case
the ``standard'' golden rule for the transition probability from one unperturbed eigenstate $\ketN{n_{\mathrm i}}$ to another (sufficiently close) one $\ketN{n_{\mathrm f}}$
by choosing $\rho(0) = \ketN{n_{\mathrm i}} \braN{n_{\mathrm i}}$ and $A = \ketN{n_{\mathrm f}} \braN{n_{\mathrm f}}$.
Indeed, the rate $\Gamma$ is then
proportional to the (ensemble-averaged) squared perturbation matrix element $\av{ \lambda^2 \lvert V_{n_{\mathrm i} n_{\mathrm f}} \rvert^2 } \simeq \lambda^2 \sigma^2(0)$ and the density of states $\varepsilon^{-1}$ as predicted by the golden rule.
Our present asymptotic analysis thus suggests that the relaxation characteristics of perturbed many-body systems can be described by Fermi's golden rule for sufficiently weak perturbations at sufficiently late times.
Crucially, we also quantify how the relaxation behavior changes beyond this regime when $\lvert g(t) \rvert^2$ deviates from the exponential form [see also, in particular, Eq.~\eqref{81} and its justification in Sec.~\ref{s43}].

The 
transition in the time domain predicted around Eq.~\eqref{80}
is illustrated in Fig.~\ref{fig:g:tc} for a step 
profile $f(x) = 0.04 \, \Theta(1.5^2 - x^2)$ [cf.\ Eq.~\eqref{10}], 
where $\Theta(x)$ denotes the Heaviside step function.
The step profile is chosen because the transition is very sharp here 
compared to ``more regular'' shapes \cite{dab20a}.

We also remark that a similar transition was observed in 
Ref.~\cite{gen13} for the response of a small system 
coupled to a bath, modeled by a related random-matrix setup 
where the perturbation profile~\eqref{5} 
was assumed 
to be given by a step function, too
(see also Sec.~\ref{s52} for a more detailed comparison 
of our work and Ref.~\cite{gen13}).
More precisely speaking,
based on a small-$t$ expansion and numerical 
evidence, the authors of Ref.~\cite{gen13} predicted
a Gaussian decay for $t \ll \Delta_v^{-1}$,
whose asymptotic behavior is consistent with~\eqref{eq:gSmallTime},
and an exponential decay with rate~\eqref{eq:gLargeTimeRate} for $t \gg \Delta_v^{-1}$.

%%%%%%%%%%%%%%%%%%%%%%%%%%%%%%%%%%%%%%%%%%%%%%%%%%%%%%%%%
\begin{figure*}
\includegraphics[scale=1]{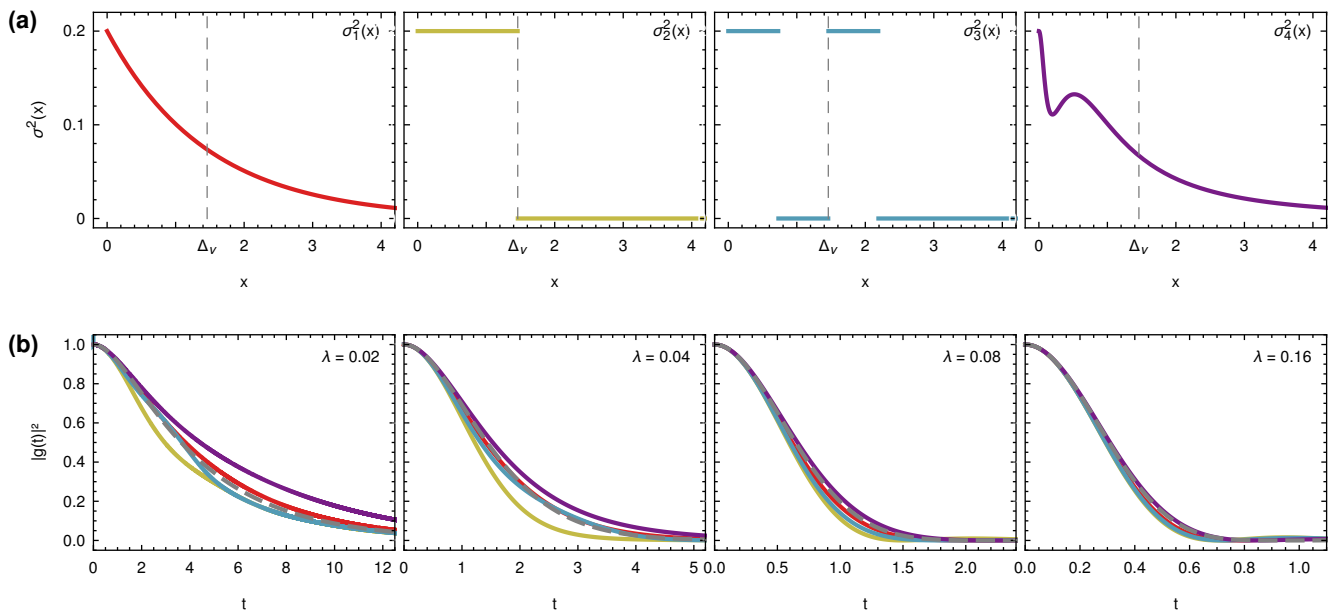}
\caption{
(a)
Four different perturbation profiles $\sigma^2(x)$, each with $\sigma^2(0) = 0.2$ and $\Delta_v = 1.5$:
exponential profile $\sigma_1^2(x) = \sigma^2(0) \, e^{-\lvert x \rvert/\Delta_v}$ 
(red),
step profile $\sigma_2^2(x) = \sigma^2(0) \, \Theta(\Delta_v^2 - x^2)$ (yellow),
two-step profile $\sigma_3^2(x) = \sigma^2(0) \, \id_{[0, \Delta_v/2] \cup [ \Delta_v, 3 \Delta_v/2]}(\lvert x \rvert)$ 
with $\id_S(x)$ denoting the indicator function of the set $S$ (blue),
sum of four Breit-Wigner functions~\eqref{5}, $\sigma_4^2(x)$, with $N = 4$, $a_1 = 2 a_2 = 4 a_3 
= 8 a_4 = 0.93$,
$2 f_1 / 3 = -f_2 = -2 f_3 = f_4 = \sigma^2(0) = 0.2$ (purple).
(b) 
Corresponding numerically exact solutions $\lvert g(t) \rvert^2$ with 
$\varepsilon = 0.002$ (hence $\alpha = 100$) for various coupling 
strengths $\lambda$ as indicated in the top-right corner of each panel
(note the different scales of the $t$-axis).
Additionally, the gray dashed line in every panel (sometimes hardly visible)
represents the analytical approximation~\eqref{81}.
}
\label{fig:gBandProfilesCompare}
\end{figure*}
%%%%%%%%%%%%%%%%%%%%%%%%%%%%%%%%%%%%%%%%%%%%%%%%%%%%%%

Combining the observations from Secs.~\ref{s41} and~\ref{s42},
we expect that the decay characteristic on the relevant time scales where $g(t)$ notably 
deviates from zero is predominantly exponential for $\lambda \lesssim \lambda_{\mathrm c}$ 
and predominantly Bessel-like for $\lambda \gtrsim \lambda_{\mathrm c}$ with 
$\lambda_{\mathrm c}$ from~\eqref{79}.
Since the relaxation time scale associated with $\lambda_{\mathrm c}$ (obtained by 
substituting $\lambda_{\mathrm c}$ into $\Gamma^{-1}$ from~\eqref{77} or 
$\gamma^{-1}$ from~\eqref{78}) is $t_{\mathrm c}/4$,
the transition in the time domain will usually only be observable in the weak-perturbation 
regime with $\lambda \lesssim \lambda_{\mathrm c}$.

%%%%%%%%%%%%%%%%%%%%%%%%%%%%%%%%%%%%%%%%%%%%%%%%%%%%%%%%%
\subsection{Refined prediction of perturbed relaxation}
\label{s43}

Next we turn to the question of how general (or special) our analytically 
considered class of functions $\sigma^2(x)$ of the 
form (\ref{10}), (\ref{12}) actually is.
To begin with we recall that the Fourier transform 
$\tilde h(k):=\int dx \, e^{-ikx}\,h(x)$ of the
function $h(x):=[1+(x/a)^2]^{-1}$ is 
$\tilde h(k)=a\pi\, e^{-a \lvert k \rvert}$.
Accordingly, the Fourier transform of $f(x)$ from (\ref{12})
is of the form $\tilde f(k)=\sum_{n=1}^N b_n\, e^{-a_n \lvert k \rvert}$
with largely general $a_n \in\RR^+$ and 
$b_n:=\pi a_n f_n\in\RR$.
In turn, from the method of Laplace transformations one
can conclude that a very large class of functions $\tilde f(k)$
can be written -- at least in very good approximation --
as sums of exponentially decaying terms.
In conclusion, the functions $f(x)$ of the form (\ref{12}) 
are actually expected to cover a very large class
of different functions $\sigma^2(x)$ in (\ref{10}).
Since $\sigma^2(x)$ is a smooth function by construction as mentioned below Eq.~\eqref{6},
it may even be expected that (\ref{12}) 
essentially covers all cases of notable interest.

In particular, the symmetry (\ref{24}) is thus 
predicted to be fulfilled very generally,
and $g(t)$ in (\ref{8}) to be even and real-valued
(see below~\eqref{28}).

Since, as pointed out above, all those perturbation profiles
$\sigma^2(x)$
with identical parameters $\alpha$ and $\Delta_v$ entail
very similar functions $g(t)$ for sufficiently small and large 
$\lambda$ (and likewise for sufficiently small and large $t$),
it is interesting to explore in more detail how much
those functions $g(t)$ will differ for intermediate
values of $\lambda$ (or of $t$).
While we did not succeed to arrive at any interesting analytical
insights along these lines, the numerical results presented
in Fig.~\ref{fig:gBandProfilesCompare} very convincingly show that the functions
$g(t)$ actually depend very little on any further details
of $\sigma^2(x)$, once the values of $\lambda$, $\alpha$, 
and $\Delta_v$ are fixed.

On the other hand, we found in Sec. \ref{s32} that
the second-order approximation (\ref{39}) reproduces
the exact behavior remarkably well for perturbation profiles $f(x)$
of the form (\ref{12}) with $N=1$. Rewriting this approximation 
(\ref{39}) in terms of the two parameters $\alpha$ and 
$\Delta_v$ according to (\ref{74})-(\ref{76}) thus
yields as a very general prediction for the
function $g(t)$ the approximation
\eqref{81} as stated in Sec.~\ref{sMainResults}.
For comparison, this approximation is also displayed 
in Fig.~\ref{fig:gBandProfilesCompare} 
as dashed gray lines.
Furthermore, even better agreement with the results for different perturbation profiles could be achieved by introducing ``renormalized'' parameters $\alpha$ and $\Delta_v$ to approximate them by the Breit-Wigner shape~\eqref{12}.

Altogether, Eqs. (\ref{7}) and (\ref{81}) thus amount to
a fully analytical prediction for the perturbed relaxation
behavior in terms of only two very basic characteristics
of the perturbation,
namely the (intrinsic) perturbation strength from (\ref{73})
and the band width (or perturbation range) from
(\ref{72}).
This is the main result of our present paper.

%%%%%%%%%%%%%%%%%%%%%%%%%%%%%%%%%%%%%%%%%%%%%%%%%%%%
%%%%%%%%%%%%%%%%%%%%%%%%%%%%%%%%%%%%%%%%%%%%%%%%%%%%
\section{Comparison with related previous works}
\label{s5}

The question of how the observable relaxation 
behavior of an unperturbed many-body 
system is modified in response to a weak 
perturbation is very natural.
Accordingly, various aspects of this general question 
have been addressed in a considerable number
of previous works.
Before discussing in somewhat more detail those
which are particularly close to our present paper,
we exemplify that even rather strong approximations
may still lead to quite decent results
in this context.

%%%%%%%%%%%%%%%%%%%%%%%%%%%%%%%%%%%%%%%%%%%%%%%%%%%%
\subsection{Simple approximation}
\label{s51}

In a first step, we employ the basis transformation
(\ref{4}) to rewrite the right hand side of
(\ref{2}) in terms of the unperturbed matrix 
elements of $\rho(0)$ and $A$,
resulting after a couple of elementary manipulations in
\begin{eqnarray}
\< A \>_{\! \rho(t)} 
\!\!\! & = & \!\!\!\!\!
\sum_{k,l,m,n} 
e^{i (E^0_n - E^0_m) t}  \rho^0_{kl}(0)  A^0_{nm} 
S^\ast_{km}(t) S_{ln}(t)
\, , \ \ \ \
\label{2a}
\\
S_{km} (t)
\!\!\! & := & \!
\sum_\mu e^{i (E_\mu-E^0_m) t} \, U^\ast_{\mu k} U_{\mu m}
\ .
\label{2b}
\end{eqnarray}
In view of the orthonormality relation
$\sum_\mu U^\ast_{\mu k} U_{\mu m}=\delta_{km}$,
it seems reasonable to expect that, within a very crude 
approximation, the sum on the right hand side of (\ref{2b})
is negligibly small unless $k=m$. 
In turn, for $k=m$ the factors $|U_{\mu m}|^2$ can be reasonably well
approximated by their ensemble-averaged values $\av{ |U_{\mu m}|^2 }$.
The main argument is that the sum consist of so many 
terms of similar character that --
much like in the common central limit theorem --
a kind of ``self-averaging'' effect can be expected.
Due to analogous arguments as above (\ref{5}), 
this average can furthermore be written in the form
\begin{eqnarray}
\av{|U_{\mu m}|^2} = u(E^0_\mu-E^0_m)
\label{2c}
\end{eqnarray}
for some suitable function $u(E)$ (see also (\ref{a5})),
and the differences $E_\mu-E^0_m$ in (\ref{2b})
and $E^0_\mu-E^0_m$ in (\ref{2c}) can be roughly
approximated by $(\mu-m)\lvsp$.
Altogether we thus arrive at 
\begin{eqnarray}
S_{km}(t) & \simeq &  \delta_{km}\,\tilde g(t)  \,,
\label{2d}
\\
\tilde g(t) & := & \sum_n e^{\I n \lvsp t} u( n \lvsp)
\ .
\label{2e}
\end{eqnarray}
Taking into account (\ref{2a}) and (\ref{3}), 
we thus can conclude that
\begin{eqnarray}
\< A \>_{\!\rho(t)} \simeq  \lvert \tilde g(t) \rvert^2 \,  \< A \>_{\!\rho_0(t)} 
\ .
\label{2f}
\end{eqnarray}

Anticipating that $\tilde g(t)$ will turn out to coincide with 
$g(t)$ (see also Eqs.~(\ref{a5}) and (\ref{a8})), 
the similarity of (\ref{2f}) and (\ref{7}) is remarkable.
If the long-time limit $\< A \>_{\!\bar\rho}$ in (\ref{7})
happens to vanish, the two relations are even identical.
The main shortcoming of our above ``quick and dirty'' derivation
is that (\ref{2f}) is clearly wrong if $A$ is the identity operator.
More generally, (\ref{7}) exhibits a wrong transformation 
behavior when adding a constant to $A$.
A more detailed inspection \cite{dab20,nat18} reveals two main
reasons for this deficiency: Though the terms in (\ref{2b})
with $k\not=m$ are indeed very small, there are
so many of them which contribute to (\ref{2a})
that they cannot be neglected.
Similarly, there are very weak but very numerous
``correlations'' between the summands in the two 
$S$-factors in (\ref{2a}), which also sum up to a 
non-negligible correction.
Yet another shortcoming is the fact that all those
arguments appear reasonable at most in cases 
where the basis transformation $U_{mn}$ from 
(\ref{4}) behaves in a ``typical'' manner, which
cannot be specified more precisely. In particular,
the likelihood to encounter an ``exception''
remains unknown.

Apart from those issues, the main point of the above
exercise is to demonstrate that at least the basic 
structure of the correct result from (\ref{7}) is remarkably
robust against quite crude approximations in its derivation.

%%%%%%%%%%%%%%%%%%%%%%%%%%%%%%%%%%%%%%%%%%%%%%%%%%%%
\subsection{Connections to pertinent previous works}
\label{s52}

The overall setting of our approach, involving a reference Hamiltonian 
$H_0$ perturbed by an ensemble of random matrices, is similar to 
Deutsch's seminal paper~\cite{deu91}.
In this early work, the considered perturbations are taken from the Gaussian 
orthogonal ensemble (GOE) and thus do not emulate the sparsity and 
bandedness properties found in many physical examples,
but the pertinent modifications are subordinate with respect to the 
questions studied in~\cite{deu91}, see also~\cite{rei15, dab20, dab20a}.
Namely, Ref.~\cite{deu91} establishes that the absence of thermalization 
in many-body quantum systems is fragile in the sense that small 
perturbations will typically lead to thermal equilibrium phenomenology 
in the long-time limit.
To do so, Deutsch devised a computational scheme
\cite{deuYY} to approximately calculate ensemble averages 
for products of eigenvector overlaps~\eqref{4} between the 
unperturbed and perturbed systems.
In particular, from the so-obtained results for the
second moments in (\ref{2c}) one can infer that
the function $\tilde g(t)$ in (\ref{2d}) indeed agrees 
with our previous $g(t)$ from (\ref{8})
for the considered special case that the perturbation
matrices are sampled from a GOE.
It is therefore no surprise that such eigenvector 
overlap moments are actually also at the heart of 
the derivations in Ref.~\cite{dab20} 
(see also Eqs. (\ref{a5}) and (\ref{a8})),
even though different methods were used there for their 
evaluation.
Also note that the focus of our present work (as well as its 
predecessors \cite{dab20, dab20a}) is on the dynamical 
relaxation process rather than the equilibrium 
(long-time average) properties.

Using Deutsch's random matrix model, and extending his 
approximative methodology, 
Nation and Porras \cite{nat19} found a result akin 
to Eq.~\eqref{7} for the time evolution of expectation values 
with the special choice $g(t) = \e^{-\Gamma t / 2}$.
Since the perturbations were again sampled from 
a GOE, whose band width in (\ref{72}) is infinite, 
this is in agreement with the special case discussed 
below (\ref{78}).
Technically speaking, their approximation includes 
some of the omitted small terms discussed below 
(\ref{2f}), while others are still missing
as a consequence of the fact that some basic 
orthonormality properties of the unitary matrix 
$U_{mn}$ are not properly accounted for 
(see equation (48) in \cite{nat18}).
As a consequence, considerable restrictions with
respect to the class of admitted observables $A$ 
had to be made in \cite{nat19},
requiring a special type of sparsity where only a few 
off-diagonals
of the matrix $A^0_{mn}$ 
[see below Eq.~\eqref{3}] may exhibit 
nonvanishing entries.
Furthermore, the concentration-of-measure property, 
crucial for turning ensemble averages into predictions 
about the overwhelming majority of individual 
members of the considered perturbation ensemble 
(see above (\ref{7})), was merely postulated 
rather than proved in Ref.~\cite{nat19}.

By means of yet another approach, based 
on a Lippman-Schwinger-type equation,
Ithier and Ascroft evaluated special cases of the 
eigenvector overlap moments in Deutsch's random 
matrix ensemble up to fourth order \cite{ith18},
which are needed in the last two factors in (\ref{2a})
when evaluating the ensemble average of that
equation.

The investigation of significantly more 
general random matrix ensembles of 
the form~\eqref{1}, covering, among others,
perturbations exhibiting a sparse and/or 
banded structure, has a long-standing history as well, see, e.g., 
Refs.~\cite{wig55,wig57,mir91,fei91,zyc92,pro93,jac95,fyo95,fyo96,cas96,shl96,sre99} 
and references therein.
Particularly noteworthy
in our present context
is the work~\cite{fyo96} by Fyodorov et al.,
who studied the statistical properties of banded and sparse random 
matrix ensembles closely related to the perturbation ensembles 
in~\cite{dab20,dab20a} and the present work.
Notably, the integral equation~\eqref{9}
is equivalent to an integral equation
which was obtained in a similar but different 
context in Ref.~\cite{fyo96}. Namely, by
means of the definition $g_\kappa(z) := i [ G(z)^{-1} - z ]$
one readily recovers from our Eq.~(\ref{9}) the 
integral equation (10) in Ref. \cite{fyo96}.
In addition, yet another, even more involved integral equation
has been obtained in \cite{fyo96} [see Eq.~(5) therein]
which has no counterpart in our present context.
A possible reason could be that the random matrix ensembles
considered in \cite{fyo96} differ from ours in some subtle details.
In any case, the derivation of both equations has not been
provided in Ref.~\cite{fyo96} nor in any subsequent work by
these authors \cite{pc}.
To our knowledge, the first published derivation 
of~(\ref{9}) is thus contained in the Supplemental 
Material of Ref.~\cite{dab20}.

Incidentally, a predecessor of 
the integral equation~\eqref{11}
can already be
 found in Wigner's seminal investigations \cite{wig55, wig57} of 
 banded perturbation matrices whose entries have constant amplitude but random signs.
In particular, with the definition $p(z) := i G(z)$, Eqs.~(9a) and~(9b) 
in \cite{wig57} correspond to our Eq.~\eqref{11} for a step profile 
$f(x) = q \, \Theta(1 - x^2)$.

A special case of the present banded random matrix model 
was adopted by Genway et al.\ in Ref.~\cite{gen13},  
investigating the particular setup of a small 
system in contact with a large heat bath,
and focusing on the temporal relaxation 
behavior of the small system.
Formally, the matrix elements $V^0_{mn}$ were introduced 
as independent, unbiased Gaussian random variables with 
variance~\eqref{5} and a step profile
$f(x) = \alpha \, \Theta(\Delta_v^2 - x^2)$,
even though the effective structure is slightly distorted 
due to the employed Dyson Brownian motion approach \cite{dys62} 
to approximately compute the eigenvector overlap moments 
which are needed in (\ref{2a}) and (\ref{2b}) \cite{gen13}.
In this setting, Genway et al.\ then obtained an approximate
prediction for the  time dependence of the subsystem's reduced 
density matrix which resembles our 
present results \eqref{7}--(\ref{9}) [being asymptotically exact, 
as explained below Eq. (\ref{7})].
Notably, they conclude that the coupling to the bath 
(which is the perturbation in the setup of their work) modifies 
the isolated reference dynamics by a characteristic damping 
function which is approximately Gaussian for times 
$t < \Delta_v^{-1}$ and crosses over to an exponential shape 
thereafter; we had already briefly commented on this 
transition in Sec.~\ref{s4} below Eq.~\eqref{80}.

Yet another approximation in terms of projection operator 
techniques is due to Richter et al. \cite{ric19}, 
adopting a quite similar (non-rigorous) line of reasoning 
as in Sec. \ref{s51}, and indeed arriving at a result of the
same formal structure as in (\ref{2f}), except that
the function $\tilde g(t)$ is now defined somewhat 
differently, and that the admitted initial states $\rho(0)$
must satisfy certain additional restrictions.

As announced in Sec. \ref{s51}, the fact that all those 
diverse approaches and approximation schemes of 
varying degree of rigor often lead to quite similar
conclusions indicates an astonishing structural stability 
of the problem and underpins its fundamental nature.
Nevertheless, 
an entirely
satisfying treatment can apparently 
not reasonably circumvent such quite sophisticated and 
technically involved calculations as those elaborated in 
Ref. \cite{dab20}.

%%%%%%%%%%%%%%%%%%%%%%%%%%%%%%%%%%%%%%%%%%%%%%%%%%%%
%%%%%%%%%%%%%%%%%%%%%%%%%%%%%%%%%%%%%%%%%%%%%%%%%%%%
\section{Comparison with numerical examples}
\label{s6}

So far, our main predictions (\ref{7}), (\ref{81})
amount to quite well controlled analytical approximations 
regarding the vast majority of all members of a given random 
matrix ensemble of perturbations.
But what are their implications regarding the ``true''
(non-random) perturbation when dealing with 
some specific physical model in (\ref{1})\,?

Similarly as when comparing ``true'' random numbers with 
numerically generated pseudo-random numbers, it is in 
general very difficult to decide whether or not the ``true'' perturbation 
$V$ in any concrete physical model can be reasonably well 
captured by such a random matrix approach.
Indeed it may seem {\em a priori} almost obvious that the 
true perturbation matrix is not a random matrix.
On the other hand, the following counter-argument is equally
obvious: Let us assume that the matrix in question
is of large but finite dimension and that each matrix element 
can in principle assume a large but finite number of different
possible values. (Such a simplification does not seem to 
entail very serious problems, since it is unavoidable 
in any numerical treatment.) Sampling each possible 
value of the matrix elements with some finite probability  
then gives rise to an ensemble of random matrices,
which also contains the ``true'' perturbation as one 
of its members.
In particular, if the ensemble imitates some key features
like sparsity, banded matrix structure etc.\ of the true
perturbation reasonably well, and if we are able
to make a statement which applies to the overwhelming
majority of all the members of the ensemble
(as in our present case), then there is good reason to
believe that this statement should also apply 
to the true perturbation (see also Introduction).
Yet, this counter-argument can again be 
countered as follows:
If the statement in question quantitatively depends 
on certain finer details of the considered ensemble, then
the ``true'' perturbation can at most belong to the
vast majority of one ensemble, while it must belong
to the tiny minority of all other ensembles.
In other words, the previous argument that the 
true perturbation is realized with some reasonable 
probability is not yet sufficient to guarantee that it belongs 
to the majority of all the similarly behaving members
of a given ensemble.
Again, this objection is mitigated in our present case
by the fact that the prediction (\ref{7}), (\ref{81})
only depends on two basic parameters of the considered
ensemble, i.e., many different ensembles effectively
behave in the same way.

As usual in random matrix theory (see also Introduction), 
the only really convincing way out
is to compare our predictions (\ref{7}), (\ref{81})
with the actual behavior of concrete examples.

%%%%%%%%%%%%%%%%%%%%%%%%%%%%%%%%%%%%%%%%%%%%%%%%%%%%
\subsection{Fermionic Hubbard model}
\label{s61}

%%%%%%%%%%%%%%%%%%%%%%%%%%%%%%%%%%%%%%%%%%%%%%%%%%%%
\begin{figure}
\includegraphics[scale=1]{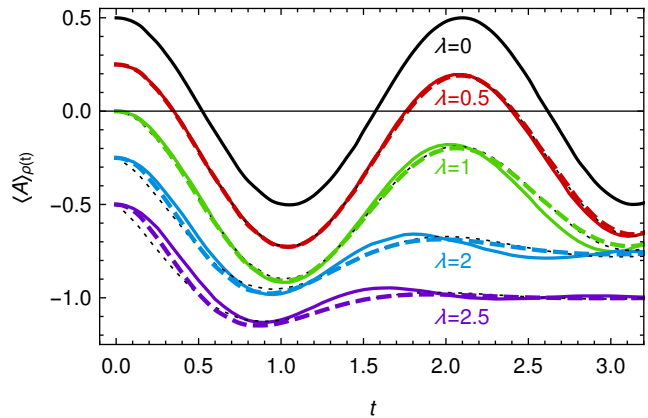}\\
\caption{
Time-dependent expectation values of the momentum mode 
correlation~\eqref{eq:ExBalzer:A} for the fermionic Hubbard
model from~\eqref{eq:ExBalzer:H} with various 
coupling strengths $\lambda$, and with a N\'eel state 
as initial condition.
Solid: DMFT results as published in Fig.~3(a) of Ref.~\cite{bal15}.
Dashed: Theoretical prediction~\eqref{7} using the numerical result 
(solid black $\lambda=0$ curve) for the unperturbed $\< A \>_{\!\rho_0(t)}$, 
$\< A \>_{\!\bar\rho} = 0$, and $g(t)$ from~\eqref{81} with 
$\alpha = 0.045$ and $\Delta_v = 4.88$.
Dotted: Same but employing for $g(t)$
the approximation from~\eqref{37} with $\alpha = 0.035$.
Data for finite $\lambda$ are shifted vertically 
in steps of $-0.25$ for better visibility.
}
\label{fig:ExBalzer}
\end{figure}
%%%%%%%%%%%%%%%%%%%%%%%%%%%%%%%%%%%%%%%%%%%%%%%%%%%%

As a first example, we consider a variant of the fermionic Hubbard 
model for which Balzer et al.\ obtained the perturbed relaxation dynamics 
numerically in Ref.~\cite{bal15}.
A comparison with the typicality prediction from 
Ref.~\cite{dab20} had already been included in the Supplemental 
Material there, and we conjectured that visible deviations for 
short times and larger perturbation strengths are caused by a banded 
structure of the perturbation matrix $V^0_{mn}$.
Accordingly, we now employ our
refined analytical prediction~\eqref{7} with 
$g(t)$ from~\eqref{81}, which explicitly accounts for effects of a 
decaying perturbation profile via the parameter $\Delta_v$, 
to further examine this conjecture.

The considered model is defined on a Bethe lattice of infinite 
coordination number so that $H_0$ and $V$ in~\eqref{1} can be written as
\begin{equation}
\label{eq:ExBalzer:H}
	H_0 := -\sum_{\<ij\>,\sigma} c^\dagger_{i\sigma} c_{j\sigma} \,,
	\quad
	V := \sum_i (n_{i\sUp} - \tfrac{1}{2}) (n_{i\sDn} - \tfrac{1}{2})
\end{equation}
using the fermionic creation and annihilation 
operators $c^\dagger_{i\sigma}$ and $c_{i\sigma}$, 
respectively, for particles of spin $\sigma \in \{ \sUp, \sDn \}$ 
on site $i$, $n_{i\sigma} := c^\dagger_{i\sigma} c_{i\sigma}$, 
and $\<ij\>$ to denote connected sites $i$ and $j$.
Initially, every site is occupied by exactly one fermion with opposing 
spins between connected sites (N\'eel state).
Time-dependent expectation values for the observable
\begin{equation}
\label{eq:ExBalzer:A}
	A := \tfrac{1}{2} \left( \hat c^\dagger_k \hat c_{\bar k} 
	+ \hat c^\dagger_{\bar k} \hat c_k \right) ,
\end{equation}
which quantifies correlations between conjugated momentum 
modes $k$ and $\bar k$ as detailed in~\cite{bal15},
were calculated for various coupling strengths $\lambda$ 
using dynamical mean-field theory (DMFT), and are displayed 
as solid lines in Fig.~\ref{fig:ExBalzer}.

In order to compare these numerical findings with our
analytical prediction~\eqref{7}, we take the numerically 
obtained reference dynamics $\<A\>_{\!\rho_0(t)}$ 
($\lambda = 0$ in Fig.~\ref{fig:ExBalzer}) and the 
long-time asymptotics $\< A \>_{\!\bar\rho} = 0$
as ``given''.
For the function $g(t)$ from~\eqref{81}, we need the 
numerical values of the parameters $\Delta_v$ from~\eqref{72} 
and $\alpha$ from~\eqref{73}.
In principle, they can be extracted directly from the empirical 
variances~\eqref{5} of the operator 
$V$ from~\eqref{eq:ExBalzer:H} 
(see Sec.~\ref{s6X} and Ref.~\cite{dab20a} for an example where this analysis is explicitly employed).
For the present system, unfortunately, these values are not 
available from Ref.~\cite{bal15}, so we treat them as fit 
parameters, yielding the dashed curves 
in Fig.~\ref{fig:ExBalzer}.
Using the corresponding best fits $\alpha = 0.045$ 
and $\Delta_v = 4.88$ one furthermore finds with 
Eqs.~(\ref{79}) and (\ref{80}) that 
$\lambda_c\simeq 4.7$ and $t_c\simeq 0.64$.

For comparison, we also show 
in Fig.~\ref{fig:ExBalzer} as dotted curves
the prediction of the simpler approximation~\eqref{37} 
for $g(t)$ with only one fit parameter $\alpha$.
This approximation was employed previously in Ref.~\cite{dab20},
and is {\em de facto} restricted -- as pointed out below (\ref{78}) 
and above (\ref{80}) -- to small coupling strengths 
$\lambda$, or large band widths $\Delta_v$, or large times $t$, 
while otherwise the two-parameter approximation
(\ref{81}) should be better.
Fig.~\ref{fig:ExBalzer} indeed confirms our theoretical
prediction that the improved approximation (dashed lines)
works notably better than the simple approximation (dotted lines)
when
$t<t_c\simeq 0.64$,
albeit there still remain some noticeable deviations from the
numerics (solid lines) for large 
$\lambda$- and moderate $t$-values (for very small and large 
$t$-values good agreement is found for rather trivial reasons).
These deviations are probably due to the fact that 
for too strong perturbations the assumption below
(\ref{3}) regarding the mean level spacing may 
easily be violated: Either the system's energy distribution 
may no longer be sufficiently sharply peaked, or the
mean level spacing itself may be notably changed 
by the perturbation.

Furthermore,
the remaining differences
between the solid and dashed lines in Fig.~\ref{fig:ExBalzer}
may possibly be attributed to finite size effects 
(see below Eq. (\ref{7})) and the effective approximation 
of the ``true'' perturbation profile~\eqref{5} by the Breit-Wigner form~\eqref{12}, 
or they may indicate that the specific model 
Hamiltonian at hand indeed exhibits some non-negligible
deviations from the vast majority of the typical 
members for any of the random matrix ensembles 
admitted by our present approach (see beginning of 
this section).

While it may not be overly surprising that a two-parameter fit performs better than a one-parameter fit to approximate the actual dynamics,
it is still worth pointing out that the improvement of the two-parameter approximation~\eqref{81} primarily concerns the regime of larger $\lambda$ and small $t$, which is precisely the region where it is anticipated to enhance the prediction (see also Secs.~\ref{s3} and~\ref{s4}).
For smaller values of $\lambda$ or larger $t$, in turn, both approximations essentially agree (cf.\ Fig.~\ref{fig:ExBalzer}).

As an aside we note that any given random 
matrix ensemble entails a unique value of the 
parameter $\alpha$ in (\ref{73}). 
The fact that the fitted $\alpha$ values notably
differ for the two above mentioned approximations 
may be considered as yet another signature of the
fact that the simpler approximation
(dotted lines) indeed misses some relevant
feature of the true system, namely the banded
matrix structure of the perturbation.

A noteworthy final remark is that in the 
above example the unperturbed expectation
value does not approach a constant long-time limit
(the unperturbed system does not equilibrate 
nor thermalize), 
but rather keeps oscillating forever, and that our
theory also admits such cases (as announced in Sec. \ref{s2}).

%%%%%%%%%%%%%%%%%%%%%%%%%%%%%%%%%%%%%%%%%%%%%%%%%%%%
\begin{figure}
\includegraphics[scale=1]{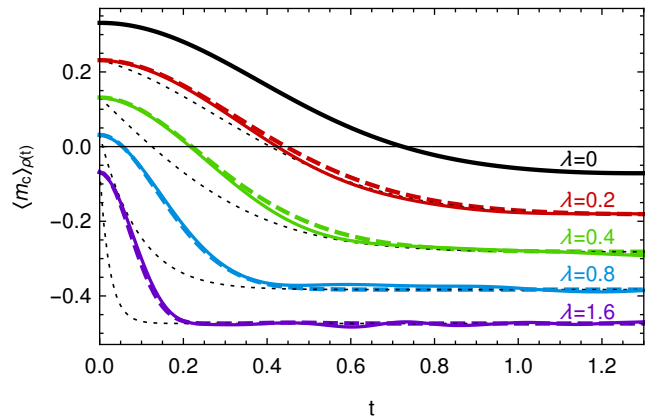}
\caption{
Time-dependent expectation values of the central magnetization correlation $A = m_{\mathrm c}$ from~\eqref{eq:ExSpin4x4:MagCorr} for the $4 \times 4$ spin-$\frac{1}{2}$ lattice model from~\eqref{eq:ExSpin4x4:H0}--\eqref{eq:ExSpin4x4:V}.
The initial state is given by $\rho(0) = \ket\psi \bra\psi$ with $\ket\psi$ 
from~\eqref{eq:ExSpin4x4:InitState}.
Solid: Numerical results obtained by exact diagonalization.
Dashed: Theoretical prediction~\eqref{7} using the numerical result 
(solid black $\lambda = 0$ curve) for the unperturbed $\< A \>_{\!\rho_0(t)}$, 
the numerically determined $\< A \>_{\!\bar\rho} = -0.0896, -0.0820, -0.0830, -0.0738$ 
for $\lambda = 0.2, 0.4, 0.8, 1.6$,
respectively, and $g(t)$ from~\eqref{81} with 
$\alpha = 2.64$ and $\Delta_v = 7.32$
[see below Eq.~\eqref{eq:ExSpin4x4:V}]. 
Dotted: Same but employing for $g(t)$ the approximation from~\eqref{37}.
Data for finite $\lambda$ are shifted vertically in steps of $-0.1$.
}
\label{fig:ExSpin4x4}
\end{figure}
%%%%%%%%%%%%%%%%%%%%%%%%%%%%%%%%%%%%%%%%%%%%%%%%%%%%

%%%%%%%%%%%%%%%%%%%%%%%%%%%%%%%%%%%%%%%%%%%%%%%%%%%%
\subsection{Two-dimensional spin-1/2 lattice}
\label{s6X}

As our second example, we consider a two-dimensional spin-$\frac{1}{2}$ lattice of dimensions $L \times L$,
in which the unperturbed Hamiltonian $H_0$ couples nearest neighbors via isotropic Heisenberg interactions (open boundary conditions),
\begin{equation}
\label{eq:ExSpin4x4:H0}
	H_0 := \sum_{i,j=1}^{L} \bm\sigma_{i,j} \cdot \left(\bm\sigma_{i+1,j} + \bm\sigma_{i, j+1} \right) \,,
\end{equation}
while the perturbation provides an additional coupling of next-nearest neighbors via spin-flip terms with respect to the $z$ direction,
\begin{equation}
\label{eq:ExSpin4x4:V}
	V := \sum_{i,j=1}^{L-1} \sum_{\alpha=x,y} \left( \sigma^\alpha_{i,j} \sigma^\alpha_{i+1,j+1} + \sigma^\alpha_{i+1,j} \sigma^\alpha_{i,j+1} \right) .
\end{equation}
Here $\bm\sigma_{i,j} := (\sigma^x_{i,j}, \sigma^y_{i,j}, \sigma^z_{i,j})$ is a vector of Pauli matrices acting on site $(i,j)$.
For a lattice with $L = 4$, the corresponding perturbation profile $\sigma^2(x)$ from~\eqref{5} was computed numerically using exact diagonalization in Ref.~\cite{dab20a}, yielding $\alpha = 2.64$ and $\Delta_v = 7.32$ within an energy window from $E = -8.8$ to $5.8$.
The associated crossover coupling from~\eqref{79} is thus $\lambda_{\mathrm c} \simeq 0.75$,
and the crossover time from~\eqref{80} is $t_{\mathrm c} \simeq 0.43$.

We prepare the system in the state $\rho(0) = \ket{\psi} \bra{\psi}$ with
\begin{equation}
\label{eq:ExSpin4x4:InitState}
	\ket\psi \propto e^{-H_0^2 / 2 \Delta^2} \sigma^+_{2,2} \sigma^+_{3,3} \ket\phi \,,
\end{equation}
where $\Delta = 2$, $\sigma^+_{i,j} := \sigma^x_{i,j} + \I \sigma^y_{i,j}$, and $\ket\phi$ denotes a Haar-distributed random vector in the Hilbert space sector with vanishing total magnetization in the $z$ direction.
This emulates a system in thermal equilibrium at infinite temperature, for which the spins at sites $(2,2)$ and $(3,3)$ are prepared in the ``up'' state, followed by a macroscopic energy measurement yielding $E = 0$.

Starting from this state, we monitor the correlation between the $z$ magnetization of the two initially deflected sites,
\begin{equation}
\label{eq:ExSpin4x4:MagCorr}
	m_{\mathrm c} := \sigma^z_{2,2} \, \sigma^z_{3,3} \,.
\end{equation}
The numerically obtained time-dependent expectation values are shown as solid lines in Fig.~\ref{fig:ExSpin4x4} for various values of the coupling strength $\lambda$,
spanning the entire regime from weak perturbations ($\lambda < \lambda_{\mathrm c}$) to moderately strong perturbations ($\lambda > \lambda_{\mathrm c}$).

To compare our analytical prediction~\eqref{7},
we adopt the function $g(t)$ from~\eqref{81} with the known values $\alpha = 2.64$ and $\Delta_v = 7.32$ [see below Eq.~\eqref{eq:ExSpin4x4:V}] and the numerically available expectation values $\<A\>_{\!\bar\rho}$ (cf.\ figure caption).
This leads to the dashed curves in Fig.~\ref{fig:ExSpin4x4}, which are in remarkable agreement with the numerical results.
Notably, there are no free fit parameters in this example,
all quantities entering the prediction \eqref{7}, \eqref{81} are available directly via a numerical analysis of $H_0$ and $V$ from~\eqref{eq:ExSpin4x4:H0} and~\eqref{eq:ExSpin4x4:V}, respectively.

In addition, we also display the prediction~\eqref{7} with the weak-perturbation and long-time asymptotics~\eqref{37} for $g(t)$ by the dotted lines in Fig.~\ref{fig:ExSpin4x4}.
This approximation indeed works in the expected regimes ($\lambda \ll \lambda_{\mathrm c}$ or $t \gg t_{\mathrm c}$),
but obviously fails for other combinations of $\lambda$ and $t$,
highlighting the substantial improvement of the theory resulting from the refined expression~\eqref{81} for $g(t)$.

%%%%%%%%%%%%%%%%%%%%%%%%%%%%%%%%%%%%%%%%%%%%%%%%%%%%
\subsection{Spin-1/2 XXX ladder}
\label{s62}

Our third example consist in a spin-$\frac{1}{2}$ ladder model,
where $H_0$ and $V$ in (\ref{1}) are given by
\begin{equation}
\label{eq:ExRichter:H}
	H_0 := \sum_{l=1}^L \sum_{k=1}^2 \bm S_{l,k} \cdot \bm S_{l+1,k} \,,
	\quad
	V := \sum_{l=1}^L \bm S_{l,1} \cdot \bm S_{l,2} \,.
\end{equation}
Here
$\bm S_{l,k} = \bm\sigma_{l,k}/2$
denote standard 
spin-$\frac{1}{2}$ operators 
acting on site $(l,k)$.
The reference system $H_0$ thus consists of two spin 
chains (or ``legs'') of length $L$ with nearest-neighbor Heisenberg 
interactions, and the perturbation $V$ couples them via 
Heisenberg terms as well.
In other words, this model exemplifies a case of two 
isolated many-body subsystems, brought into contact by 
the perturbation $V$ (see Sec.~\ref{s2}).

Richter et al.\ studied in Ref. \cite{ric19}
the current autocorrelation function 
$C(t) := \tr\{ \rho_{\mathrm{eq}} e^{i H t} J e^{-i H t} J \}/L$ 
of this model, where $\rho_{\mathrm{eq}}$ 
denotes the canonical density operator at infinite temperature 
(thus proportional to the identity operator), 
and where
\begin{equation}
\label{eq:ExRichter:A}
	J := \sum_{l=1}^L \sum_{k=1}^2 \left( S^x_{l,k} S^y_{l+1,k} - S^y_{l,k} S^x_{l+1,k} \right)
\end{equation}
is the spin current along the legs.
Within the dynamical typicality framework they utilized
to simulate $C(t)$ \cite{ric19}, and exploiting the fact that $\tr\{J\} = 0$, 
the autocorrelation function $C(t)$ from above
can be rewritten in the form $\< J \>_{\!\rho(t)}$ (cf.\ Eq.~(\ref{2}))
with initial state $\rho(0) = \ket{\psi}\bra{\psi}$,
where $\ket{\psi}$ is chosen proportional to $\sqrt{\id + J/L} \, \ket{\phi}$
\cite{f3},
and $\ket{\phi}$ being a state drawn at random from 
the pertinent Hilbert space according to the Haar 
measure \cite{ric19}.

%%%%%%%%%%%%%%%%%%%%%%%%%%%%%%%%%%%%%%%%%%%%%%%%%%%%
\begin{figure}
\includegraphics[scale=1]{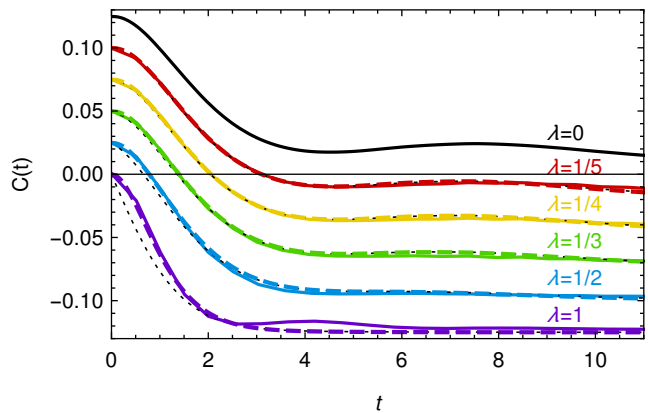}\\
\caption{
Time-dependent expectation values of the current correlation function 
$C(t) = \tr[ \rho_{\mathrm{eq}} e^{i H t} J e^{-i H t} J ]/L = \< J \>_{\!\rho(t)}$ 
(see also main text) for the spin ladder model~\eqref{eq:ExRichter:H}.
Solid: Numerical results as provided in Fig.~2(c) of Ref.~\cite{ric19}.
Dashed: Theoretical prediction~\eqref{7} using the numerical result 
(solid black $\lambda=0$ curve) for the unperturbed $\< J \>_{\!\rho_0(t)}$, 
$\< J \>_{\!\bar\rho} = 0$, and $g(t)$ from~\eqref{81} 
with $\alpha = 0.13$ and $\Delta_v = 3.26$.
Dotted: Same but employing for $g(t)$
the approximation from~\eqref{37} with $\alpha = 0.11$.
Data for finite $\lambda$ are shifted vertically in steps of $-0.025$ 
for better visibility.
}
\label{fig:ExRichter}
\end{figure}
%%%%%%%%%%%%%%%%%%%%%%%%%%%%%%%%%%%%%%%%%%%%%%%%%%%%

The numerical results for $C(t) =  \<J\>_{\!\rho(t)}$ from 
Ref.~\cite{ric19} are depicted as solid lines in Fig.~\ref{fig:ExRichter}.
We compare them to our theory~\eqref{7},~\eqref{81}
by exploiting $\< J \>_{\!\bar\rho} = 0$ 
(thermal long-time asymptotics)
and fitting the parameters $\alpha$ and $\Delta_v$ since 
-- as in the previous example from Sec.~\ref{s61} --
their precise numerical values are not available.
(We remark that the authors of~\cite{ric19} do explore 
the structure of their perturbation matrix $V^0_{mn}$,
but for a smaller system size
than in Fig. \ref{fig:ExRichter},
and indeed find a banded and sparse structure 
in qualitative agreement with our setup.
Unfortunately, the precise scaling of the parameters $\alpha$ 
and $\Delta_v$ with the system size $L$ is unclear, 
so that it is not possible to 
extrapolate their quantitative values for the system
in Fig.~\ref{fig:ExRichter} from the available data.)
The dashed lines in Fig.~\ref{fig:ExRichter} represent the 
so-obtained theoretical predictions with $\alpha = 0.13$ and $\Delta_v = 3.3$,
revealing very good agreement with the numerics.
Similarly as in the previous example~from Sec.~\ref{s61}, we also included as dotted 
lines a fit to the simpler approximation from~\eqref{37}, 
which is expected (and found) to apply when the perturbation is weak, 
or the band width is large, or the time is large.

%%%%%%%%%%%%%%%%%%%%%%%%%%%%%%%%%%%%%%%%%%%%%%%%%%%%
\subsection{Transverse-field Ising model}
\label{s63}

\begin{figure*}
\includegraphics[scale=1]{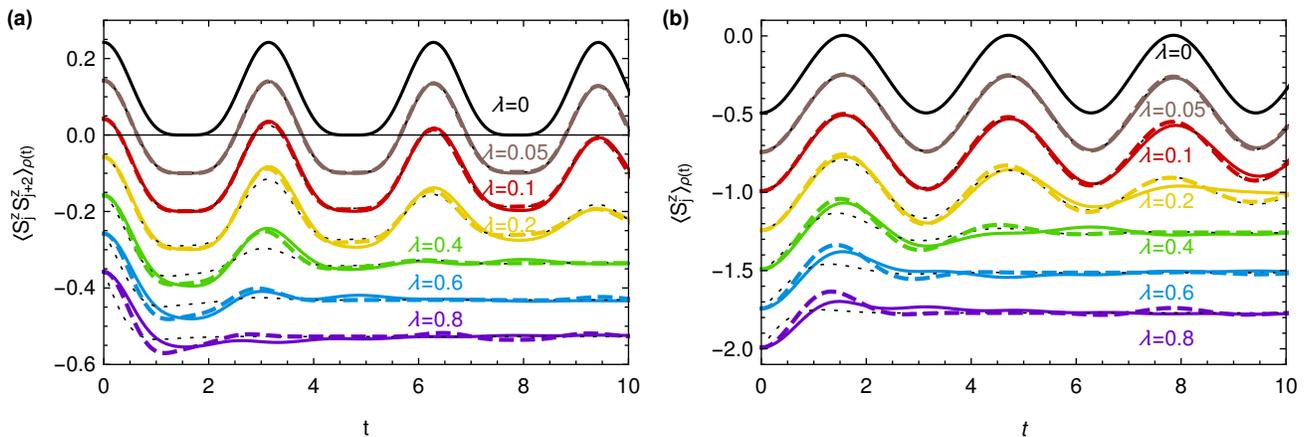}\\
\caption{
Time-dependent expectation values of 
(a) the next-nearest-neighbor correlation $A = S^z_j S^z_{j+2}$ and 
(b) the transverse magnetization $A = S^z_j$ for the transverse-field 
Ising model~\eqref{eq:TFIM:H} with various coupling
strengths $\lambda$, and employing the ground state 
for $\lambda=4$ as initial condition (quantum quench).
Solid: Exact analytical results in the thermodynamic limit $L\to\infty$.
Dashed: Theoretical prediction~\eqref{7} using the exact result 
(solid black $\lambda=0$ curve)
for the unperturbed $\< A \>_{\!\rho_0(t)}$, 
the exact long-time limit $\< A \>_{\!\bar\rho}$, and $g(t)$ 
from~\eqref{81} with $\alpha = 1.1$ and $\Delta_v = 0.22$.
[In particular, the same values were use for all dashed 
curves in (a) and (b).]
Dotted: Same but employing for $g(t)$
the approximation from~\eqref{37} with $\alpha = 0.70$.
Data for finite $\lambda$ are shifted vertically in steps of 
$-0.1$ in (a) and $-0.25$ in (b) for better visibility.
}
\label{fig:ExTFIM}
\end{figure*}

As a last example, we consider the one-dimensional transverse-field 
Ising model, whose Hamiltonian is given by (\ref{1}) with
\begin{equation}
\label{eq:TFIM:H}
	H_0 := -2 \sum_{j=1}^L S^x_j S^x_{j+1} \,,
	\quad
	V := -\sum_{j=1}^L S^z_j
	\ ,
\end{equation}
where the $S^{x,z}_j$ are again the standard 
spin-$\frac{1}{2}$ operators 
acting on site $j$ modulo $L$ (periodic boundary conditions).
Hence, the reference system is the Ising model at vanishing 
external field, and the perturbation consists of a transverse 
magnetic field with coupling strength $\lambda$.
This model is well-known to be
integrable for all values of $\lambda$.

Specifically, we consider an infinite chain
(thermodynamic limit $L\to\infty$), 
which is prepared in the ground state 
of the perturbed Hamiltonian with $\lambda = 4$, 
and which is subsequently quenched to a Hamiltonian with a smaller 
$\lambda$ value.
{As our first observable, we consider the next-nearest neighbor 
correlation $S^z_j S^z_{j+2}$ (the choice of $j$ is irrelevant), 
which is known to be nonthermalizing in the present setup \cite{vid16}.
Our second observable is the transverse magnetization $S^z_j$.
Both observables exhibit permanent oscillations in the reference system 
$H_0$ for the chosen initial state, and the expectation-value dynamics 
can be calculated exactly for all values of $\lambda$ 
\cite{igl00, dzi05, pus16}.

Similarly as before, we compare in Fig.~\ref{fig:ExTFIM}
those
numerically
exact results (solid lines) to our theory from \eqref{7} and \eqref{81}, 
employing the exact results also for the reference 
dynamics $\< A \>_{\!\rho_0(t)}$ 
and for the long-time limit $\< A \>_{\!\bar\rho}$, 
while $\Delta_v$ and $\alpha$ are again treated as fit parameters,
yielding $\Delta_v = 0.22$ and $\alpha = 1.2$.
The improved approximation 
(dashed lines) again works notably better than 
the simple approximation (dotted lines) when 
$\lambda > \lambda_c\simeq 0.2$.

We also note that -- like in the example from Sec.~\ref{s61} -- we are dealing 
here again  with a case for which the unperturbed dynamics 
does not equilibrate in the long-time limit.
Moreover, both the unperturbed and the perturbed systems
are now always integrable.
Though this does not automatically violate any of the formal requirements (cf.\ Sec.~\ref{s2}),
our theory
may not have been originally expected to be still applicable
in view of the fact that most perturbations of any given ensemble will lead to a non-integrable perturbed Hamiltonian.
For the particular combination of observables and initial state from Fig.~\ref{fig:ExTFIM},
the presence of an extensive number of conservation laws thus seems to be of subleading importance as far as the dynamical response of the system to the perturbation is concerned.

%%%%%%%%%%%%%%%%%%%%%%%%%%%%%%%%%%%%%%%%%%%%%%%%%%%%
%%%%%%%%%%%%%%%%%%%%%%%%%%%%%%%%%%%%%%%%%%%%%%%%%%%%
\section{Conclusions}
\label{s7}

Characterizing the effect of reasonably small perturbations on the behavior 
of dynamical systems is an important and recurrent task in many areas of physics.
The pertinent key contribution of our present work is an analytical prediction of 
this response in isolated
many-body quantum systems based on just two parameters 
of the perturbation:
its overall strength $\alpha \lambda^2$ [Eq.~\eqref{73}]
and its band width $\Delta_v$ [Eq.~\eqref{72}], both of which are 
derived from the magnitude of the perturbation matrix elements 
in the eigenbasis of the unperturbed Hamiltonian via~\eqref{5}.
The analytical prediction~\eqref{7} suggests that the perturbed time-dependent 
expectation values $\< A \>_{\!\rho(t)}$ of an observable $A$ resemble the 
unperturbed $\< A \>_{\!\rho_0(t)}$, but are modified by a function $g(t)$, 
specified in~\eqref{81} and depending on the two perturbation parameters 
$\alpha \lambda^2$ and $\Delta_v$, which pushes them towards their (perturbed) 
long-time limit. 
For late times and/or weak perturbations, these perturbed relaxation characteristics are akin to Fermi's golden rule and in fact comprise the latter as a special case.
At the same time, significant deviations from the golden-rule behavior are observed and quantified via~\eqref{81} for stronger perturbations and short times.

Given its simplicity,
the prediction~\eqref{7}, \eqref{81} can surely not be expected to hold 
always and invariably.
Naturally, deviations are unsurprising if certain prerequisites 
of the derivation itself
are violated,
the most important ones being a well-defined macroscopic energy and a 
perturbation that is not 
too strong.
However, deviations can also occur in a more subtle manner due to the 
inherent probabilistic character of our random matrix approach.
Moreover, as explained in Sec.~\ref{s6}, it is unfortunately quite hard 
to determine in general whether a given physical perturbation is a ``typical'' 
member of any of the admitted perturbation ensembles.

In any case, a crucial prerequisite is that the chosen ensemble faithfully reproduces the key mechanisms responsible for the modifications in the perturbed system.
Here, we identified the perturbation profile $\sigma^2(x)$ from~\eqref{5} as the key property for the dynamics in a large variety of setups, prominently highlighted by the integro-differential equation~\eqref{eq:gIntEq}.
Nevertheless, other features such as the locality and few-body character of interactions may become important in other setups, too (see below).

Generally speaking,
it is not unusual that the behavior of macroscopic 
systems can be characterized in terms of just a few parameters, for 
otherwise we would not be able to describe them theoretically at all.
Crucially, the examples from Sec.~\ref{s6} (see also Ref.~\cite{dab20} 
for further examples) demonstrate that the theory is successful at predicting 
the dynamical response for a variety of different models and perturbations,
and even for systems which cannot really be considered ``macroscopic'' yet.

Methodologically, our present work builds on the results from Ref.~\cite{dab20},
which, in particular, established the general structure~\eqref{7} for 
the quantum many-body relaxation under the influence of banded and 
sparse perturbation matrices.
Substantial progress was achieved regarding analytical expressions for 
the function $g(t)$ governing that perturbed relaxation,
culminating in the aforementioned approximation~\eqref{81} in terms of 
the perturbation strength $\alpha \lambda^2$ and band width $\Delta_v$.
This latter expression originally arises as the second-order approximation 
for the special choice of a single ($N=1$) Breit-Wigner perturbation 
profile~\eqref{12},
for which we were able to solve the underlying integral equation~\eqref{11} 
leading to $g(t)$ via~\eqref{8} exactly in the form of an infinite continued 
fraction (see Sec.~\ref{s3}).
Based on universal limiting expressions for weak and strong perturbations 
and short and long times as well as numerical explorations, we subsequently 
argued in Sec.~\ref{s4} that the precise form of $g(t)$ is rather insensitive to 
further perturbation details apart from $\alpha \lambda^2$ and $\Delta_v$, 
finally leading to~\eqref{81}.

Besides our own works \cite{dab20, dab20a},
similar concepts and ideas have been developed 
and pursued in various studies by other authors, too,
of which we reviewed the most closely related ones 
known to us in Sec.~\ref{s5}.

An interesting direction for future research is to obtain a 
better understanding of the features of ``real'' physical 
perturbations which may potentially render them atypical
with respect to the random matrix ensembles considered 
here.
From a mathematical point of view, these are correlations 
between the perturbation matrix elements,
but how they arise from the geometrical structure and types 
of interactions \cite{nic19, sug20} and to what extent they 
matter with respect to the many-body relaxation is still poorly 
understood.

%%%%%%%%%%%%%%%%%%%%%%%%%%%%%%%%%%%%%%%%%%%%%%%%%%%%

\begin{acknowledgments}

This work was supported by the 
Deutsche Forschungsgemeinschaft (DFG)
within the Research Unit FOR 2692
under Grant No. 397303734,
by the Paderborn Center for Parallel 
Computing (PC$^2$) within the Project 
HPC-PRF-UBI2,
and by the International Centre for Theoretical Sciences 
(ICTS) during a visit for the program -  
Thermalization, Many body localization and Hydrodynamics 
(Code: ICTS/hydrodynamics2019/11).

\end{acknowledgments}

%%%%%%%%%%%%%%%%%%%%%%%%%%%%%%%%%%%%%%%%%%%%%%%%%%%%
%%%%%%%%%%%%%%%%%%%%%%%%%%%%%%%%%%%%%%%%%%%%%%%%%%%%
\appendix
%
%%%%%%%%%%%%%%%%%%%%%%%%%%%%%%%%%%%%%%%%%%%%%%%%%%%%%%%
\section{General properties of $G(z)$}
\label{app1}

In this appendix we discuss several general issues regarding 
the solutions $G(z)$ of the non-linear integral equation (\ref{9}),
including existence, uniqueness, symmetries, and the
physical origin
of this equation.

Generally speaking, statements about the existence and 
uniqueness of solutions of non-linear integral 
equations as exemplified 
by (\ref{9}) are known to be difficult to obtain.

Before turning to our specific case at hand, 
it may be worthwhile to recall the following basic facts:
The function $f(z):=z^\ast$ is not analytic 
for any $z\in\CC$ (the Cauchy-Riemann differential equations 
are always violated). It follows that if $f(z)$ is analytic, 
then $h(z):=f(z^\ast)$ is in general no longer analytic,
and similarly for $h(z):=[f(z)]^\ast$.
On the other hand, $h(z):=-f(z)$ and $h(z):=f(-z)$ are still 
analytic, and also $h(z):=[f(z^\ast)]^\ast$ is again analytic,
and hence also $h(z):=-[f(-z^\ast)]^\ast$.

Due to the symmetry (\ref{6}) of $\sigma^2(x)$, one readily verifies 
that if $G(z)$ solves the integral equation (\ref{9}), then there 
must exist in full generality four solutions of (\ref{9}), 
namely (see also (\ref{25}))
\begin{eqnarray}
G_1(z) & := & G(z) = v(x,y) + i w(x,y)
\ ,
\label{a1}
\\
G_2(z) & := & -[G(-z^\ast)]^\ast = -v(-x,y) + i w(-x,y)
\, , \ \ \
\label{a2}
\\
G_3(z) & := & [G(z^\ast)]^\ast = v(x,-y) - i w(x,-y)
\ ,
\label{a3}
\\
G_4(z) & := & -G(-z) = -v(-x,-y) - i w(-x,-y)
\, . \ \ \ 
\label{a4}
\end{eqnarray}

In general,
solutions of the non-linear integral equation (\ref{9}) 
thus come in quadruples of the structure 
(\ref{a1})-(\ref{a4}).
However, some of the members of a quadruple
may actually coincide, for instance due to 
symmetry reasons (see below).
Moreover, we anticipate that they may 
exhibit poles (or other singularities), 
and hence may not be well-defined 
or may not solve~\eqref{9}
for all $z \in \CC$ simultaneously.
In the following, we always tacitly focus on one
particular such quadruple of
solutions.

If $G(z)=G_1(z)$ furthermore satisfies the two additional
requirements (i) and (ii) below (\ref{13}), then one readily verifies
also $G_2(z)$ will satisfy the same requirements.
On the other hand, $G_3(z)$ and $G_4(z)$ will 
fulfill two slightly modified requirements, 
namely the same requirements as below (\ref{13})
except that they must now apply for all $z$ with 
$\mbox{Im}(z)>0$ 
(but possibly not any more for all $z\in\CC$ with $\mbox{Im}(z) < 0$).

Under the assumption that 
the 
solution 
of (\ref{9}),
in conjunction with the two requirements below (\ref{13}),
is unique, it follows that $G_2(z)=G_1(z)$, and hence
$G(z)=G_1(z)$ must satisfy the symmetry property 
(\ref{24}).
Incidentally, also the two remaining solutions 
$G_3(z)$ and $G_4(z)$ then
actually must coincide and exhibit the same symmetry (\ref{24}).

Further insight regarding those existence and 
uniqueness issues seems hardly possible on the 
basis of the integral equation (\ref{9}) alone.
Rather, it seems indispensable to also
take into account the origin of this equation
in the derivation of the final result (\ref{7}).
Referring to \cite{dab20} for more details, 
a suitable starting point is the following 
implicit definition of the function $u(E)$:
\begin{eqnarray}
\av{|U_{mn}|^2} = u(E^0_m-E^0_n)
\ ,
\label{a5}
\end{eqnarray}
where, as in the main paper, $U_{mn}$ are the
overlaps of the perturbed and unperturbed
eigenstates from (\ref{4}) and where $\av{\cdots}$
indicates the average over the considered ensemble
of perturbations $V$.
For the rest, everything is very similar to the implicit
definition of the function $\sigma^2(x)$ in (\ref{5}).
The existence and uniqueness of this function $u(E)$ 
may thus be taken for granted.
Moreover, by means of our usual approximation
$E^0_m - E^0_n\simeq (m-n)\lvsp$ (see above (\ref{5})),
the following two properties readily follow from the 
definition (\ref{a5}):
\begin{eqnarray}
u(E)\geq 0
\ ,
\label{a6}
\\
\int dE\, u(E) = \lvsp 
\ .
\label{a7}
\end{eqnarray}

The key point of the derivation of the final result (\ref{7})
in Ref. \cite{dab20} is the following relation between the 
functions $u(E)$ and $G(z)$:
\begin{eqnarray}
u(E) = \frac{\varepsilon}{\pi}\lim_{\eta\downarrow0} \,\mbox{Im}\,G(E-i\eta)
\ ,
\label{a8}
\end{eqnarray}
where $G(z)$ must solve the non-linear integral equation (\ref{9}).
On the other hand, $G(z)$ originally arises \cite{dab20}
as an ensemble average of the form $\av{ \mathcal G(z)  } = G(z - H_0)$, 
where $\mathcal G(z) := (z - H)^{-1}$ 
is the resolvent or Green's function
of the perturbed Hamiltonian $H$ in (\ref{1}),
see also below~\eqref{8}.
For the sake of clarity, we temporarily denote this specific
(original) function as $\tilde G(z)$, to better distinguish it from
all the possibly existing further solutions of the non-linear 
integral equation (\ref{9}).
Taking for granted that the entire derivation in \cite{dab20}
is sound, it follows that this ensemble-averaged (scalar)
resolvent $\tilde G(z)$ exists, is unique, and solves (\ref{9}).
Moreover, one can infer from the above definitions
that the Hermiticity of $\mathcal G(z)$ implies
$\tilde G(z^\ast)=[\tilde G(z)]^\ast$, 
and that $\tilde G(z)$ must in general be 
expected to exhibit non-analyticities along 
the real axis  \cite{dab20}.
In contrast, the (exact or approximate) solutions
we encountered in the main paper usually do not 
exhibit the symmetry $G(z^\ast)=[G(z)]^\ast$,
and instead are often analytic on the real axis.
Put differently, the essential difference between 
our solutions $G(z)$ and the original $\tilde G(z)$ 
is that $G(z)$ initially was obtained in the main
paper as a solution of the non-linear integral
equation (\ref{9}) on the complex half-plane $\CC^-$
from (\ref{13}), and subsequently was extended
as far as possible (i.e. up to poles) by analytic
continuation.
On the other hand, $\tilde G(z)$ may be imagined to
arise in the same way as a solution of  (\ref{9}) 
on the half-plane $\CC^-$, but then is already
fixed also for the remaining arguments $z$ 
via $\tilde G(z^\ast)=[\tilde G(z)]^\ast$.
In other words, if $\tilde G(z)$ agrees with $G(z)$ in (\ref{a1})
on $\CC^-$, then $\tilde G(z)$ will be given by 
$G_3(z)$ from (\ref{a3}) for the remaining arguments $z$.
[In particular, $\tilde G(z)$ thus still solves the non-linear
integral equation (\ref{9}) for all 
$z\in\CC\setminus\RR$.]

Let us now assume that we obtained a quadruple of
solutions (\ref{a1})-(\ref{a4}), and one of them, say $G_1(z)$,
leads via (\ref{a8}) to a function $u(E)$ which satisfies 
(\ref{a6}) and (\ref{a7}).
Then the same applies for the corresponding 
function $u(E)$ deriving from $G_2(z)$.
On the other hand, $G_3(z)$ and $G_4(z)$ will entail
functions $u(E)$ via (\ref{a8}) which violate (\ref{a6})
and (\ref{a7})
(or they are not well-defined for the arguments 
$z$ required in (\ref{a8}) in the first place).
Hence, one of the first two solutions is expected to be 
the physically relevant one (i.e. to agree with 
$\tilde G(z)$ on $\CC^-$), while the last two 
solutions must be excluded as unphysical.

All these arguments strongly suggest (but still do
not prove), that a solution $G(z)$ of the non-linear
integral equation (\ref{9}), for which
$u(E)$ from (\ref{a8}) moreover satisfies
(\ref{a6}) and (\ref{a7}), exists and is unique.
Furthermore, the uniqueness strongly suggests
that also the symmetry property (\ref{24})
will be fulfilled.
According to (\ref{25}), (\ref{27}), and (\ref{a8}) 
it then also follows that $u(E)$ must 
be an even function of $E$,
\begin{eqnarray}
u(-E)=u(E)
\ .
\label{a9}
\end{eqnarray}

All these properties were indeed recovered
in all our particular (possibly approximate
and either analytical or numerical)
solutions $G(z)$ from the main paper.
[Note that (\ref{a7}) is in fact equivalent to (\ref{35}) 
due to (\ref{a8}) and (\ref{8}).]

%%%%%%%%%%%%%%%%%%%%%%%%%%%%%%%%%%%%%%%%%%%%%%%%%%%%%%%
\section{Convergence of Eq.~\eqref{23}}
\label{app2}

In this appendix, we show that the infinite continued-fraction 
representation of $H(z)$ from Eq.~\eqref{23} converges for 
all $z \in \mathbb{C}$ with $\operatorname{Re}(z) > 0$
and any given $\beta\in\RR^+$.
Introducing the notation
\begin{equation}
\label{eq:ContFrac}
	\CFK_{k=0}^\infty \frac{a_k}{b_k}
		:= \frac{ a_0 }{ b_0 + \frac{ a_1 }{ b_1 + \frac{ a_2 }{ b_2 + \cdots } } }
\end{equation}
for the infinite continued fraction with coefficients 
$a_k, b_k \in \mathbb{C}$, we can express $H(z)$ in~\eqref{23} as
\begin{equation}
\label{eq:HCF}
	H(z) = \frac{1}{\beta} \CFK_{k=0}^\infty \frac{\beta}{z + k} \,.
\end{equation}
We also define the auxiliary continued fractions
\begin{equation}
\label{eq:hCF}
	h_n(z) := \CFK_{k=n}^\infty \frac{\beta}{z + k} \,.
\end{equation}
Provided that $h_{n+1}(z)$ converges for given $n$, Eq.~\eqref{eq:HCF} 
can then be written as
\begin{equation}
\label{eq:HCFSplitted}
	H(z) % = \frac{1}{\beta} \frac{ \beta }{ z + \frac{ \beta }{ z + 1 + \frac{ \beta }{ \;\;\ddots \frac{ \beta }{ z + n + h_{n+1}(z) } } } }
		= \frac{1}{\beta} \CFK_{k=0}^{n} \frac{\beta}{z + k + \delta_{nk} h_{n+1}(z) } \,,
\end{equation}
i.e., as a finite continued fraction of order $n$ with the ``rest'' $h_{n+1}(z)$ 
appended to the denominator of the last term.

Restricting ourselves to $\operatorname{Re}(z) > 0$ and recalling that $\beta > 0$,
it therefore suffices to show that, for fixed $z$, there exists $n_0 \in \mathbb{N}$ 
such that $h_{n_0}(z)$ is convergent
because the denominators of all finite fractions in~\eqref{eq:HCFSplitted} have 
a strictly positive real part and thus cannot develop poles.

The existence of such $n_0$ for any preset $z = x + i y$ with $x>0$ and $y \in \mathbb{R}$ 
follows directly from the \'Slezy\'nski-Pringsheim theorem \cite{lor92}, which states that 
the infinite continued fraction~\eqref{eq:hCF} converges if $\lvert z + k \rvert \geq \beta + 1$ 
for all $k \geq n$.
If $\lvert z \rvert \geq \beta + 1$, we can thus readily conclude that $h_n(z)$ converges 
for all $n$.
Otherwise,
the theorem assures convergence of $h_n(z)$ for all $n \geq n_0$ with
\begin{equation}
	n_0 := \left\lceil \sqrt{ (\beta+1)^2 - y^2 } - x \right\rceil ,
\end{equation}
where $\lceil x \rceil$ denotes the smallest integer greater than or equal to $x$ (``ceiling function'').

%%%%%%%%%%%%%%%%%%%%%%%%%%%%%%%%%%%%%%%%%%%%%%%%%%%%%%%
\section{Various representations of $g(t)$}
\label{app3}

In this appendix, we derive several alternative equations 
for the function $g(t)$ from (\ref{8}) under the assumption 
that $G(z)$ fulfills the requirements (i) and (ii) below (\ref{13}).
In particular, the integral equation~\eqref{eq:gIntEq}
for $g(t)$ is obtained.
Moreover, it is shown that $g(t)$ can be written in 
the form (\ref{28}) provided $G(z)$ exhibits 
the symmetry property (\ref{24}).

Representing $\mbox{Im}\, G(E-i\eta)$ as
$[G(E-i\eta)-G^\ast(E-i\eta)]/2i]$, the function
$g(t)$ from (\ref{8}) can be rewritten in the form
\begin{eqnarray}
g(t) & = & \lim_{\eta\downarrow 0} g_{\eta}(t)
\ ,
\label{c1}
\\
g_{\eta}(t) & := &
\frac{\tilde g_{\eta}(t)-\tilde g^\ast_{\eta}(-t)}{2\pi i}
\label{c2}
\\
\tilde g_{\eta}(t) & := & \int dE\, e^{iEt}\, G(E-i\eta)
\ .
\label{c3}
\end{eqnarray}
In particular, taking for granted that the integral
on the right hand side of (\ref{8}) exists, also
the existence of the Fourier transform (\ref{c3})
should not be a problem \cite{f2}.

From (\ref{c1}) and (\ref{c2}) we can conclude that
\begin{eqnarray}
g(-t)=g^\ast (t)
\ ,
\label{c4}
\end{eqnarray}
and with (\ref{a7}), (\ref{a8}) that
\begin{eqnarray}
g(0)=1
\ .
\label{c5}
\end{eqnarray}
In doing so, we thus tacitly take for granted
the relations (\ref{a7}) and (\ref{a8}).
Alternatively, (\ref{c5}) is also recovered
as the limit of $g(t)$ for asymptotically small $t$
in all specific examples known to us, see
also Eq. (\ref{35}) in the main text.

In view of (\ref{c4}) and (\ref{c5}),
we can and will mostly focus on $t>0$ from now on.

Taking for granted that $G(z)$ fulfills the assumptions 
(i) and (ii) below (\ref{13}),
standard residue methods imply that 
$\tilde g_{\eta}(t)$ in (\ref{c3})
must vanish for $t<0$ implying with (\ref{c2}) that
\begin{eqnarray}
g_{\eta}(t) 
= 
\tilde g_{\eta}(t)/2\pi i
\ \ \mbox{for $t>0$}
\ .
\label{c6}
\end{eqnarray}
With (\ref{c1}) and (\ref{c3}), we thus obtain
\begin{eqnarray}
g(t) & = & \frac{1}{2\pi i}\,  \lim_{\eta\downarrow0}
\int dx\, G(x-i\eta) \, e^{ix t }
\ \ \mbox{for $t>0$}
\ .
\label{c7}
\end{eqnarray}

If $G(z)$ in addition satisfies the symmetry 
property (\ref{24}), one can infer from (\ref{c3}) 
that $\tilde g^\ast_{\eta}(t)=-\tilde g_{\eta}(t)$.
With (\ref{c1}) and (\ref{c2}) it follows that
$g(-t)=g(t)$ and with (\ref{c4}) that
$g(t)=g^\ast(t)$ for all $t$.
Exploiting (\ref{c7}) we thus recover (\ref{28}),
at least for all $t\not=0$.
In the remaining case $t=0$, 
(\ref{28}) follows by continuation
or from (\ref{a7}) and (\ref{a8}),
see also the discussion below (\ref{c5}).

In the following we no longer require that 
$G(z)$ exhibits the symmetry property (\ref{24}).
Next, we observe that the integral equation (\ref{11}) 
can be rewritten by choosing $z=E-i\eta$ and
setting $h(E):=G(E-i\eta)$ in the form
\begin{eqnarray}
[E-i\eta]\,  h(E) - h(E)\int dx\, h(E-x)f(x) = 1 
\ .
\label{c8}
\end{eqnarray}
Upon multiplying this equation by $e^{iEt}$, integrating over $E$,
exploiting that $\int dE e^{iEt}h(E)=\tilde g_\eta(t)$ according to
(\ref{c3}), and applying textbook Fourier transformation methods,
a straightforward but somewhat tedious calculation yields
\begin{equation}
\frac{\dot{\tilde g}_\eta(t)}{i} - i\eta \tilde g_\eta(t) -
\int \frac{ds}{2\pi} \, \tilde g_\eta(t - s) \, \tilde g_\eta(s) \, \tilde f(s) 
= 2\pi \delta(t)
\ ,
\label{c9}
\end{equation}
where $\tilde f(t)$ is the Fourier transform of $f(x)$
from~\eqref{eq:fFT}.
Focusing on $t>0$ and utilizing under the integral in (\ref{c8})
that $\tilde g_{\eta}(t)=0$ for $t<0$ (see above (\ref{c6})), 
we can conclude that
\begin{equation}
\dot{\tilde g}_\eta(t) + \eta \tilde g_\eta(t) +
\int_0^t \frac{ds}{2\pi i} \, \tilde g_\eta(t - s) \, \tilde g_\eta(s) \, \tilde f(s) 
= 0 
\label{c11}
\end{equation}
for $t>0$.
With (\ref{c6}) and (\ref{c1}) this finally yields
\begin{equation}
\dot  g(t) +  \int_0^t ds \, g(t - s) \, g(s) \, \tilde f(s) 
= 0 
%\ \ \ \mbox{for $t>0$,}
\label{c12}
\end{equation}
for $t>0$.
Formally, this amounts to an initial value problem for
$g(t)$ with initial condition (\ref{c5}). The solution for
negative $t$ then follows from (\ref{c4}),
and can be shown to still satisfy (\ref{c12}).
In other words, (\ref{c12}) in fact applies for
arbitrary $t$, i.e., we recover Eq.~\eqref{eq:gIntEq} 
from the main text.

%%%%%%%%%%%%%%%%%%%%%%%%%%%%%%%%%%%%%%%%%%%%%%%%%%%%%%%

%%%%%%%%%%%%%%%%%%%%%%%%%%%%%%%%%%%%%%%%%%%%%%%%%%%
%%%%%%%%%%%%%%%%%%%%%%%%%%%%%%%%%%%%%%%%%%%%%%%%%%%%


\begin{thebibliography}{70}

\bibitem{haa10}
F. Haake,
{\em Quantum Signatures of Chaos}
(Springer, Berlin, 2010).
%,Third Edition).

\bibitem{kam71}
N. G. van Kampen, 
{The case against linear response theory}. 
{Phys. Norv.}  {\bf 5}, 279
 (1971).

\bibitem{bro81}
T. A. Brody, J. Flores, J. B. French, P. A. Mello, A. Pandey, and
S. S. M. Wong, 
{ Random-matrix physics:
spectrum and strength fluctuations},
{Rev. Mod. Phys.}  {\bf 53}, 385 (1981).

\bibitem{dab20}
L. Dabelow and P. Reimann,
Relaxation Theory for Perturbed Many-Body Quantum Systems versus Numerics and Experiment,
Phys. Rev. Lett. {\bf 124}, 120602 (2020).

\bibitem{deu91}
J. M. Deutsch, 
{ Quantum statistical mechanics in a closed system},
{ Phys. Rev. A } {\bf 43}, 2046 (1991).

\bibitem{gen13}
S. Genway, A. F. Ho, and D. K. K. Lee,
Dynamics of thermalization and decoherence of a nanoscale system,
Phys. Rev. Lett. {\bf 111}, 130408 (2013).

\bibitem{nat19}
C. Nation and D. Porras,
Quantum chaotic fluctuation-dissipation theorem: 
effective Brownian motion in closed quantum systems,
Phys. Rev. E {\bf 99}, 052139 (2019).

\bibitem{gol10a}
S. Goldstein, J. L. Lebowitz, R. Tumulka,  and  N. Zangh\`{\i},
{ Long-Time Behavior of Macroscopic Quantum Systems: 
Commentary Accompanying the English Translation of 
John von Neumann's 1929 Article on the Quantum 
Ergodic Theorem},
{Eur. Phys. J. H}  {\bf 35}, 173 (2010).

\bibitem{gol10b}
S. Goldstein, J. L. Lebowitz, C. Mastrodonato,  R. Tumulka, and N. Zangh\`{\i},
{ On the Approach to Thermal Equilibrium of Macroscropic Quantum Systems},
{Phys. Rev.  E}  {\bf 81}, 011109 (2010).

\bibitem{gog16}
C. Gogolin and J. Eisert,
Equilibration, thermalization, and the emergence
of statistical mechanics in closed quantum systems,
Rep. Prog. Phys. {\bf 79}, 056001 (2016).

\bibitem{dal16}
L. D'Alessio,  Y. Kafri, A. Polkovnikov, and M. Rigol,
{ From Quantum Chaos and Eigenstate Thermalization
to Statistical Mechanics and Thermodynamics},
{Adv. Phys.}  {\bf 65}, 239 (2016).

\bibitem{mor18}
T. Mori, T. N. Ikeda, E. Kaminishi, and M. Ueda,
{ Thermalization and prethermalization 
in isolated quantum systems: a theoretical overview},
{ J. Phys. B}  {\bf 51}, 112001 (2018).

\bibitem{bor16}
F. Borgonovi, F. M. Izrailev, L. F. Santos, and V. G. Zelevinsky, 
{ Quantum chaos and thermalization in isolated systems of 
interacting particles},
{Phys. Rep.}  {\bf 626}, 1 (2016).

\bibitem{nan15}
R. Nandkishore and D. A. Huse, 
{ Many-body localization and thermalization 
in quantum statistical mechanics},
{Annu. Rev. Condens. Matter Phys.}  {\bf 6}, 15 (2015).

\bibitem{lan16}
T. Langen,  T. Gasenzer, and  J. Schmiedmayer,
{ Prethermalization and universal dynamics in 
near-integrable quantum systems},
{J. Stat. Mech.} {\bf 2016}, 064009 (2016).

\bibitem{tas16}
 H. Tasaki,
{ Typicality of Thermal Equilibrium and
Thermalization in Isolated Macroscopic 
Quantum Systems},
{ J. Stat Phys.} {\bf 163}, 937 (2016).


\bibitem{rei15}
P. Reimann, 
Eigenstate thermalization: Deutsch's approach and beyond,
New J. Phys.  {\bf 17}, 055025 (2015).

\bibitem{nat18}
C. Nation and  D. Porras,
{ Off-diagonal observable elements from random 
matrix theory: distributions, fluctuations, and 
eigenstate thermalization},
{New. J. Phys.}  {\bf 20}, 103003 (2018).

\bibitem{dab20a}
L. Dabelow, P. Vorndamme, and P. Reimann,
Modification of quantum many-body relaxation by  perturbations exhibiting a banded matrix structure,
Phys.~Rev.~Research {\bf 2}, 033210 (2020).

\bibitem{f1}
To recover this conclusion within our present formalism,
we note that a large band width in (\ref{75}) is tantamount 
to large $a_n$-values (see also (\ref{12})).
On the other hand, the small-$\beta$ asymptotics in 
(\ref{55})-(\ref{70}) requires that the $\tilde a_n$'s
are kept fixed while the $\beta_n$'s must become small.
This is achieved by rewriting (\ref{53}) and (\ref{54})
as $a_n=a \tilde a_n$ and 
$\beta_n=\tilde f_n\tilde a_n/a$,
keeping the $\tilde a_n$ (and also the $\tilde f_n$) fixed,
and letting $a$ grow.
Hence, the band width in (\ref{75}) now indeed
grows linearly with $a$, while $\beta$ in (\ref{76}) 
decreases as $1/a$ (see also (\ref{20})).

\bibitem{sak94}
J. J. Sakurai,
{\em Modern Quantum Mechanics}
(Addison-Wesley, Reading, MA, 1994).

\bibitem{mal19thermal}
K. Mallayya, M. Rigol, and W. De Roeck,
Prethermalization and Thermalization in 
Isolated Quantum Systems,
Phys. Rev. X {\bf 9}, 021027 (2019).

\bibitem{deuYY}
J.~M.~Deutsch,
{ A closed quantum system giving ergodicity},
\url{https://deutsch.physics.ucsc.edu/pdf/quantumstat.pdf}
(unpublished).

\bibitem{ith18}
G. Ithier and S. Ascroft, 
{ Statistical diagonalization of a random biased Hamiltonian: the case of the eigenvectors},
{J. Phys. A: Math. Theo.} {\bf 51}, 48LT01 (2018).

\bibitem{wig55}
E. P. Wigner, 
{ Characteristic vectors of bordered matrices 
with infinite dimensions},
{Ann. Math.}  {\bf 62}, 548 (1955).

\bibitem{wig57}
E. P. Wigner,
Characteristic Vectors of Bordered Matrices with Infinite Dimensions II,
Ann. Math. {\bf 65}, 203 (1957).

\bibitem{mir91}
A. D. Mirlin and Y. V. Fyodorov,
Universality of level correlation function of sparse random matrices,
J. Phys. A: Math. Gen. {\bf 24}, 2273 (1991).

\bibitem{fei91}
M. Wilkinson, M. Feingold, and D. M. Leitner, 
Localization and spectral statistics in banded random matrix ensembles,
J. Phys. A: Math. Gen. {\bf 24}, 175 (1991).

\bibitem{zyc92}
K. \.Zyczkowski, M. Levenstein, M. Ku\'s, and F. Izrailev,
Eigenvector statistics of random band matrices,
Phys. Rev. A {\bf 45}, 811 (1992).

\bibitem{pro93}
T. Prosen and M. Robnik,
Energy level statistics and localization in sparsed banded random matrix ensemble,
J. Phys. A: Math. Gen. {\bf 26}, 1105 (1993).

\bibitem{jac95}
P. Jacquod and D. L. Shepelyansky,
Hidden Breit-Wigner Distribution and Other Properties of Random Matrices with Preferential Basis,
Phys. Rev. Lett. {\bf 75}, 3501 (1995).

\bibitem{fyo95}
Y. V. Fyodorov and A. D. Mirlin,
Statistical properties of random banded matrices with strongly fluctuating diagonal elements,
Phys. Rev. B {\bf 52}, R11580 (1995).

\bibitem{cas96}
G. Casati,  B. V. Chirikov,  I. Guarneri, and F. M. Izrailev, 
{ Quantum ergodicity and localization in 
conservative systems: 
the Wigner band random matrix model},
{Phys. Lett. A} {\bf 223}, 430 (1996).

\bibitem{shl96}
D. Shlyakhtenko,
Random Gaussian Band Matrices and Freeness with Amalgamation,
Int. Math. Research Notices {\bf 1996}, 1013 (1996).

\bibitem{fyo96}
Y. V. Fyodorov, O. A. Chubykalo, F. M. Izrailev, and G. Casati, 
{ Wigner random banded matrices with sparse structure: local spectral density of states},
{ Phys. Rev. Lett.}  {\bf 76}, 1603 (1996).

\bibitem{sre99}
M. Srednicki,
The approach to thermal equilibrium in quantum 
chaotic systems,
{J. Phys. A} {\bf 32}, 1163 (1999).

\bibitem{pc}
Y. V. Fyodorov, private communication.

\bibitem{dys62}
F. J. Dyson,
A Brownian-Motion Model for the Eigenvalues of a Random Matrix,
J. Math. Phys. {\bf 3}, 1191 (1962).

\bibitem{ric19}
J. Richter, F. Jin, L. Knipschild, H. De Raedt, K. Michielsen, J. Gemmer, and R. Steinigeweg,
{Exponential damping induced by random and realistic perturbations},
Phys. Rev. E {\bf 101}, 062133 (2020).

\bibitem{bal15}
K. Balzer, F. A. Wolf, I. P. McCulloch, P. Werner,  and M. Eckstein, 
{ Nonthermal melting of N\'eel order in the Hubbard model},
{ Phys. Rev. X}  {\bf 5}, 031039 (2015).

\bibitem{f3}
Note that $\id + J/L$ is a positive operator, so that its square root is well-defined.

\bibitem{vid16}
L. Vidmar and M. Rigol,
Generalized Gibbs ensemble in integrable lattice models,
J. Stat. Mech. {\bf 2016}, 064007 (2016).

\bibitem{igl00}
F. Igl\'oi and  H. Rieger,
{Long-Range Correlations in the Nonequilibrium Quantum Relaxation of a Spin Chain},
{Phys. Rev. Lett.} {\bf 85}, 3233 (2000).

\bibitem{dzi05}
J. Dziarmaga,
{Dynamics of a Quantum Phase Transition: Exact Solution of the Quantum Ising Model},
{ Phys. Rev. Lett.} {\bf 95}, 245701 (2005).

\bibitem{pus16}
T. Pu\v{s}karov  and D. Schuricht,
{Time evolution during and after finite-time quantum quenches in the transverse-field Ising chain},
{SciPost Phys.} {\bf 1}, 003 (2016).

\bibitem{nic19}
D. Nickelsen and M. Kastner,
Modelling equilibration of local many-body quantum systems by random graph ensembles,
Quantum {\bf 4}, 273 (2020).

\bibitem{sug20}
S. Sugimoto, R. Hamazaki, and M. Ueda,
Test of Eigenstate Thermalization Hypothesis Based on Local Random Matrix Theory,
arXiv:2005.06379 (2020).

\bibitem{lor92}
L. Lorentzen and H. Waadeland,
{\em Continued Fractions with Applications}
(Elsevier North-Holland, Amsterdam, 1992).

\bibitem{f2}
In the case $t=0$, the condition (ii) below (\ref{13}) 
may in general not be sufficient to guarantee the 
existence of the integral in (\ref{c3}).
[For instance, $|z \, G(z)|\to 0$ for $|z|\to \infty$ instead of (ii) 
would be sufficient. However, $G(z)$ often does not satisfy 
such a condition, as exemplified by the approximations (\ref{33})
and (\ref{64}).]
Hence, the existence must be verified on a case by 
case basis.
Alternatively, $t=0$ may be tacitly excluded in (\ref{c3}),
and $g(0)$ is then given by (\ref{c5}) rather than by (\ref{c1}),
see also the discussion below (\ref{c5}).
% 


\end{thebibliography}
\end{document}